\documentclass[reprint,showpacs,preprintnumbers,amsmath,amssymb,a4paper,nofootinbib,superscriptaddress,floatfix,aps,rmp,showkeys]{revtex4-1}

\setcitestyle{authoryear, round, semicolon, open={(}, close={)}, aysep={, }}

\DeclareRobustCommand{\baselinestretch{1.0}}

\usepackage{lineno}
\usepackage{blindtext, color}

\usepackage{graphicx}

\usepackage{dcolumn}
\usepackage{bm}
\usepackage{amsmath}
\usepackage{todonotes}
\usepackage{placeins} 

\usepackage{color}
\usepackage{colortbl}

\usepackage{txfonts}
\usepackage[utf8]{inputenc}

\usepackage{xcolor}
\usepackage[colorlinks=true, pdfstartview=FitV, linkcolor=blue, citecolor=blue, urlcolor=blue]{hyperref}

\makeatother

\begin{document}

\preprint{$\today$}

\title{Fast Pixelated Detectors in Scanning Transmission Electron Microscopy. Part I: Data Acquisition, Live Processing and Storage}

\author{Magnus Nord}
\email{Magnus.Nord@ntnu.no}
 \altaffiliation{Current address: Department of Physics, Norwegian University of Science and Technology, Trondheim 7491, Norway.}
 \affiliation{SUPA, School of Physics and Astronomy, University of Glasgow, Glasgow G12 8QQ, UK.}
 \affiliation{EMAT, Department of Physics, University of Antwerp, Antwerp 2000, Belgium.}
\author{Robert W. H. Webster}
 \affiliation{SUPA, School of Physics and Astronomy, University of Glasgow, Glasgow G12 8QQ, UK.}
\author{Kirsty A. Paton}
 \affiliation{SUPA, School of Physics and Astronomy, University of Glasgow, Glasgow G12 8QQ, UK.}
\author{Stephen McVitie}
 \affiliation{SUPA, School of Physics and Astronomy, University of Glasgow, Glasgow G12 8QQ, UK.}
\author{Damien McGrouther}
 \affiliation{SUPA, School of Physics and Astronomy, University of Glasgow, Glasgow G12 8QQ, UK.}
\author{Ian MacLaren}
 \affiliation{SUPA, School of Physics and Astronomy, University of Glasgow, Glasgow G12 8QQ, UK.}
\author{Gary W. Paterson}
 \email{Dr.Gary.Paterson@gmail.com}
 \affiliation{SUPA, School of Physics and Astronomy, University of Glasgow, Glasgow G12 8QQ, UK.}

\date{\today}

\begin{abstract}
The use of fast pixelated detectors and direct electron detection technology is revolutionising many aspects of scanning transmission electron microscopy (STEM).
The widespread adoption of these new technologies is impeded by the technical challenges associated with them.
These include issues related to hardware control, and the acquisition, real-time processing and visualisation, and storage of data from such detectors.
We discuss these problems and present software solutions for them, with a view to making the benefits of new detectors in the context of STEM more accessible.
Throughout, we provide examples of the application of the technologies presented, using data from a Medipix3 direct electron detector.
Most of our software is available under an open source licence, permitting transparency of the implemented algorithms, and allowing the community to freely use and further improve upon them.
\end{abstract}


\maketitle

\section{Introduction}

Several technological advances have been critical in the development of the scanning transmission electron microscope (STEM) from its inception \cite{von_Ardenne_1938_tem_theory} to its current status as one of the most important techniques for high resolution imaging of materials. 
Specifically, improved vacuum systems, field emission sources, aberration correction and the introduction of annular dark field (ADF) detectors \cite{Crewe1966_science, Crewe_1968_JAP_stem} were all crucial developments.
ADF detectors are typically formed of one or more PN diode segments or scintillator photomultiplier tube arrangements.
These are placed in the far field of the objective lens and sample an angular range of the diffraction pattern of the area of the sample illuminated by the electron beam.
Such devices are well-suited for use in STEM due to their fast readout, and imaging with pixel dwell times measured in microseconds is normal.
ADF imaging was initially understood as being based on Z-contrast \cite{crewe_1970_QRevBiophys, Crewe1970_highResBio}, though understanding of the contrast mechanism evolved over time \cite{Donald1979_pma}, in turn influencing the design of such detectors.
In particular, later contributions demonstrated that the inner angle of ADF detectors had to be relatively high to exclude coherent diffraction from dominating the signal \cite{Pennycook1991_um, HARTEL_1996_UltraMicros_haadf}.
Other refinements of this arrangement have been introduced over the years, including the use of split detectors for differential phase contrast \cite{Dekkers1977_dpc, Chapman1978_dpc, chapman_1990_mdpc, mcgrouther2014development}, multiple annular detectors \cite{Shibata2010_detector, Shibata2017_nature_atom_electric_fields}, and the use of bright field or annular bright field imaging \cite{HAMMEL_1995_UltraMicros_ABF, LeBeau_2009_PRB_BF, FINDLAY_2010_UltraMicros_ABF, MacLaren_2015_APLmat_haadf_bf}.
However, all these detector configurations integrate over large angular ranges of the back focal plane, resulting in the loss of most of the information contained in the diffraction pattern.
Furthermore, space constraints in the microscope's camera chamber can limit which detectors can be used simultaneously in an individual experiment, so that data acquisition may have to be repeated several times from the same area using different detectors to collect all the signals of interest.
This can lead to difficulties in correlating the information contained in images acquired in successive experiments due to drift, and results in a higher overall dose to the sample, which is undesirable for beam-sensitive samples.

Recently, building upon technologies developed for particle physics \cite{Wermes_2005_HPD_mono, Turala_2005_SI_tracking, Turchetta2007_maps, Delpierre_2014_hybrid_pix_dev}, pixelated detectors developed for X-ray imaging have been adopted for electron imaging \cite{DirectDetectorsEM, DEDComparison, mcgrouther_medipix_stem_2015, tate_2016_mandm_empad, MIR_2017_medipix3_characterisation, Tinti_2018_eiger}.
Compared to charge coupled device (CCD) based detectors, these direct electron detectors (DEDs) typically offer much lower noise levels, improved detector quantum efficiency (DQE) and modulation transfer function (MTF), some degree of radiation hardness, and crucially, fast readout of the images.
This allows the efficient recording of the entire diffraction pattern at each scan position using millisecond or sub-millisecond dwell times, enabling either improvements in or the use of different imaging modes, such as nanobeam STEM diffraction \cite{knut_muller_nbed_2019}, position averaged convergent beam electron diffraction \cite{LEBEAU_2010_um_pacbed, Ophus_APL_2017_pacbed}, atomically resolved electrostatic field mapping \cite{Hachtel2018, Fang2019}, improved magnetic \cite{matus_pixelated_stem_magnetic_2016} induction characterisation, determination of crystal periodicity along the beam direction \cite{Nord_2018_holz}, fluctuation electron microscopy (FEM) \cite{banerjee_fem_2017}, and ptychography \cite{pennycook_ptychography_2015, Yang_nature_2015_ptychography}. 
Additionally, work is in progress to also improve scanning precession electron diffraction (SPED) \cite{Rauch_SPED_2010} using such DEDs, especially because of their better noise performance that optically-coupled CCDs \cite{MacLaren_MandM_2020_sped}.
Indeed, pixelated detectors are increasingly regarded as `universal' detectors \cite{Yang_nature_2015_ptychography, tate_2016_mandm_empad, Hachtel2018, Fang2019, ophus_mm_2019_4dstem} capable of imaging under multiple modes.

Along with the many advantages that fast pixelated detectors bring, many practical limitations arise from their use, such as the ability to get real-time information from the data stream produced from a scan to enable navigation and identification of relevant sample features, and the storage and processing of very large datasets, often much larger than the available computer memory.
In this paper (Part~I), we present solutions for the hardware control, data acquisition, real-time processing and visualisation, and storage of data from fast pixelated detectors.
The majority of the software solutions presented in this work are made available under the free and open source GPLv3 licence, allowing transparency of the implemented algorithms, and the ability for anyone to use and to further improve upon them.
The names of the software packages, modules, classes and functions we present are given in \texttt{typewriter} font.

Most of the libraries reported here are implemented in Python.
Python, being an open and free programming language is rapidly becoming the standard language for many aspects of scientific computing \cite{oliphant_07_python, Gouillart_2016_python_scikit}.
In addition to its comparative ease of use, which lowers the barrier for people to contribute and minimises developer time, Python has an extensive standard library and a large ecosystem of external libraries, including ones for optimised numerical \cite{numpy} and scientific \cite{scipy} computing, image processing \cite{scikit_image}, data visualisation \cite{matplotlib}, and work flow documentation \cite{Kluyver_2016_jupyter}.
Furthermore, it is straightforward to link Python to low level C-code, allowing development of optimised routines or use of external libraries \cite{cython}.

Within the electron microscopy community, a number of Python packages have also been developed.
One example of this is HyperSpy \cite{hyperspy_library}, which contains functionality for processing data from a wide range of TEM techniques: electron energy loss spectroscopy, energy-dispersive X-ray spectroscopy, electron holography, and more standard imaging.
It also serves as a base for several other packages, such as pyXem for analysing SPED data \cite{pyxem}, Atomap for processing atomic resolution STEM data \cite{Nord2017_atomap}, and \texttt{pixStem} for working with data from fast pixelated STEM detectors \cite{pixstem}.
Several other packages exist, like rigidRegistration for doing rigid image registration of atomic resolution image stacks \cite{rigid_registration}, and wrappers for doing STEM simulations, like PyPrismatic \cite{prismatic}.
Other packages for processing data from fast pixelated STEM detectors include py4DSTEM \cite{py4dstem}, LiberTEM \cite{libertem}, pycroscopy \cite{pycroscopy}, and \texttt{fpd} \cite{fpd}.

The post-acquisition visualisation and processing of data from fast pixelated detectors using the \texttt{fpd} and \texttt{pixStem}\footnote{After submitting this paper, it was decided to merge \texttt{pixStem} with pyXem \cite{pyxem}. All of the features detailed in part~I and II of this series of papers that are related to \texttt{pixStem} are in the processes of being added to pyXem and will continue to be available. The features of the \texttt{fpd} library remain unaffected.} libraries for the structural characterisation of materials will be reported in Part~II of this work \cite{methods_part2_arxiv}.
A third and final part (yet too be submitted) will cover aspects related to differential phase contrast analysis.
Throughout all parts, we provide examples using data from a Medipix3 detector \cite{Ballabriga_2013_Medipix3RX}.
Although some sections of the codebase are specific to the use of this detector, such as aspects relating to data acquisition, many of the issues discussed and the techniques and tools presented in this part are applicable to a wide range of other detectors, while the data processing described in the forthcoming parts~II and III of this paper series are applicable to data from any detector.

This paper is organised as follows.
In Section~\ref{sec:detector}, the Medipix3 detector is briefly introduced.
Methodologies for acquiring data from it are discussed in Section~\ref{sec:data_acquisition}.
In Section~\ref{sec:live_processing}, an architecture developed to process a live data stream from a fast pixelated detector is outlined.
In Section~\ref{sec:data_storage}, the issues around data storage are discussed and our implementation is presented.
The source data and scripts to analyse the data and produce the results presented here are publicly available \cite{paper_dataset_zenodo}.

\section{Medipix3 Detector}
\label{sec:detector}
\begin{figure*}[hbt!]
  \centering
      \includegraphics[width=15cm]{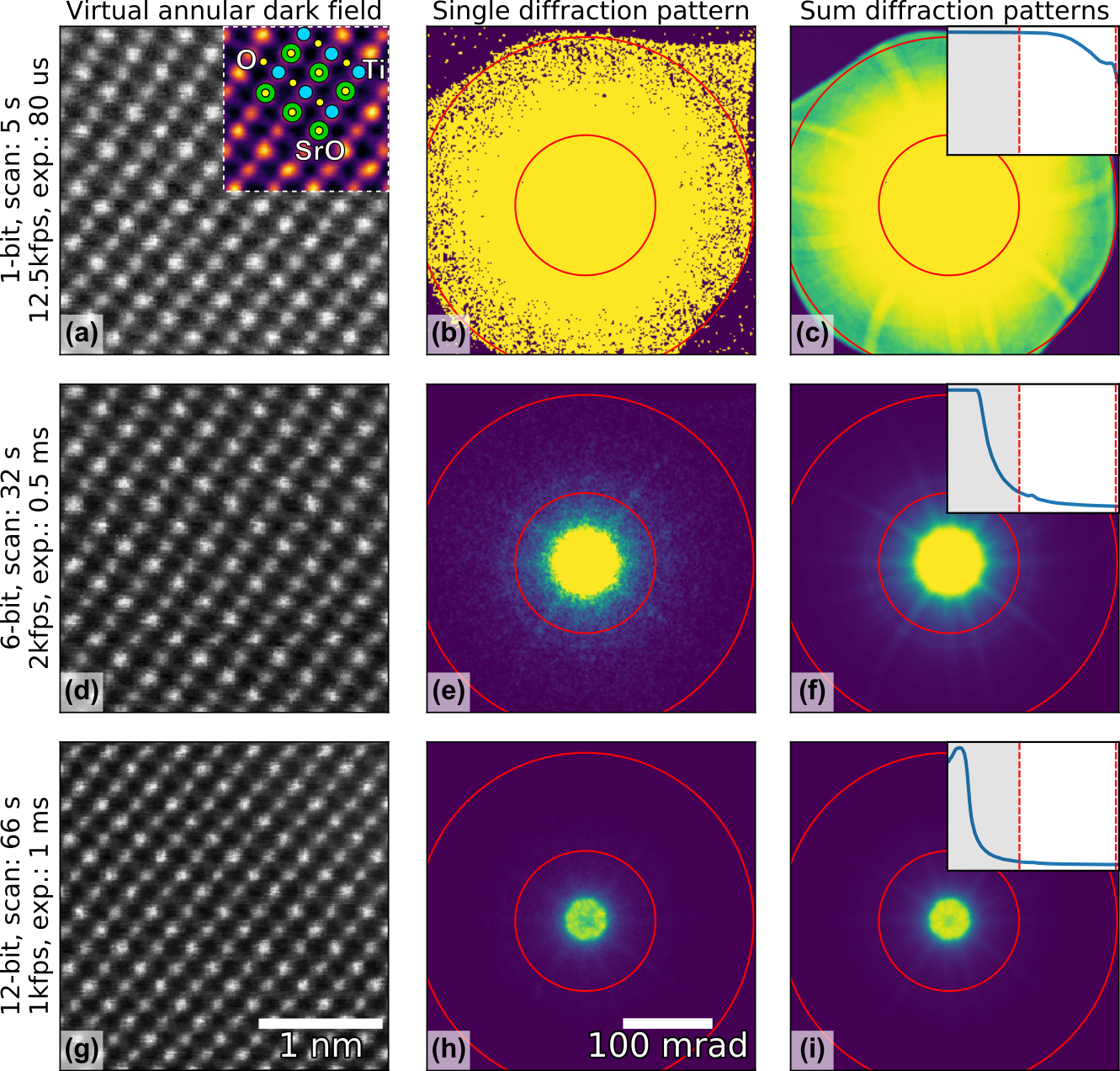}
      \caption{
          Imaging of SrTiO$_3$ along the [110] direction using different bit depths and probe dwell times (in rows) with the Medipix3 detector in continuous read-write mode.
          The bit depth, total scan time, and the detector frame acquisition rate and exposure are shown in the annotations.
          High angle ADF (HAADF) images (left column) were calculated by summing all counts inside a virtual aperture defined over the collection angles 80-192 mrad (assuming a linear mapping of pixel count to diffracted angle, which may not be entirely true in an image corrected microscope), shown by the red lines in the diffraction images (middle column), using the \texttt{pixStem} library \cite{pixstem}.
          The coloured section of the 1-bit HAADF image in (a) is Fourier filtered with a schematic overlay of the atomic columns imaged: green: Sr, yellow: O, and blue: Ti.
          The third column shows the summed diffraction patterns, with the insets displaying their radial distributions from 0 to 192~mrad.
          The dip in intensity in the centre of the direct spot in the 12-bit mode data in (h) and (i) is due to the higher bit depth now allowing the details of the primary beam and low order diffraction discs to be seen.}
      \label{fig:bit_depth_atomic_resolution}
\end{figure*}

All pixelated data reported in this work is from a 256$\times$256 pixel Medipix3RX (henceforth referred to as Medipix3) detector \cite{Ballabriga_2013_Medipix3RX} affixed to a Merlin 1R retractable Medipix3 mount from Quantum Detectors (Harwell, Oxfordshire, UK).
The Medipix3 detector is a radiation-hard hybrid counting direct electron detector, where active analogue and digital signal processing circuitry in each 55~$\muup$m pixel is bump-bonded to a relatively thick sensor layer.
Si sensor layers of 500~$\muup$m are needed for operation at primary electron energies of 300~keV.
In our case, a 300~$\muup$m silicon sensor layer was used for all data except that in Fig.~\ref{fig:bit_depth_atomic_resolution}, where a 500~$\muup$m layer was used instead.

In electron microscopy applications, an incident electron produces electron-hole pairs in the sensor layer in sufficient numbers \cite{scholze_mean_1998} for the signal due to a primary electron to be clearly distinguishable from noise in the detector.
This makes the detector capable of noiseless operation by the setting of an appropriate threshold for counting, and the detector is thus able to detect individual electrons.
As a consequence, the Medipix3 detector is of potential use in time resolved electron microscopy experiments, where sub-100~ns time resolution has been recently demonstrated \cite{medipix3_time_resolution}.

Each pixel can operate independently, with its active circuitry processing only the signal induced in that pixel, in a mode of operation known as single pixel mode (SPM).
Alternatively, in so-called charge summing mode (CSM), neighbouring pixels can pool their circuitry and collectively process the signals induced in each pixel \cite{Ballabriga_2013_Medipix3RX}.
CSM attempts to account for charge spread between pixels due to electron-matter interactions in the thick sensor layer.
At an acceleration voltage of up to 80~kV, the Medipix3 has a near-perfect DQE and MTF when imaging electrons \cite{MIR_2017_medipix3_characterisation}.
The use of alternative high-Z sensor layer materials is expected to improve the performance at higher acceleration voltages \cite{MCMULLAN_UM_2007_mpx_high_z_comment} and is currently being investigated.

Another notable feature of the Medipix3 detector is the ability to operate in continuous read-write mode, where one of the two sets of counters in each pixel is used to readout the data while the other takes over counting.
This gapless recording maximises dose efficiency, which is important for beam sensitive samples, and also enables faster acquisitions, which is important for minimising artefacts due to microscope instabilities, particularly when imaging with atomic resolution.

The Medipix3 detector can be operated in 1, 6, 12, and 24 bit depth modes, allowing the compromise between readout time, file size and dynamic range to be varied.
The clock on the Medipix3 was designed to be driven at frequencies up to 200~MHz but, with additional cooling, it can be overclocked to allow faster operation.
With the 120~MHz clock rate of the Merlin readout system \cite{Plackett_2013_merlin} used here, the readout times are 70.8~$\muup$s, 412~$\muup$s, 822~$\muup$s and 1.64~ms, for 1, 6, 12, and 24 bit modes, respectively.
While the 24 bit mode is ideal for very high dynamic range diffraction studies \cite{MIR_2017_medipix3_characterisation}, the higher readout rates of the lower bit depth modes are more generally useful across a wide range of imaging conditions (it would take $>$4~ms to exceed 12 bits at 1~MHz count rates per pixel, so 24 bits is only needed for long counting times or high arrival rates on some pixels).

To demonstrate the use of different bit depths, atomic resolution data from SrTiO$_3$ imaged along the [110] direction was acquired on a Medipix3 detector at bit depths of 1, 6 and 12, giving maximum counts of 1, 63, and 4095, respectively.
The data was acquired on a JEOL ARM 300CF using an acceleration voltage of 200~kV and a convergence angle of 22.4~mrad, with the Medipix3 operated in SPM with continuous read-write enabled.
High angle ADF (HAADF) images produced from these datasets are shown in the left hand column of Fig.~\ref{fig:bit_depth_atomic_resolution}, with the bit depth increasing from top, (a) to bottom, (g).
The atomic resolution contrast in these images arises mostly from incoherent scattering of the electrons, similar to that in regular HAADF imaging with dedicated annular detectors.
The middle and right hand columns show individual and summed diffraction patterns from each scan, respectively.
The circular red lines in the diffraction patterns mark the edges of the virtual aperture used within which pixel counts were added up to give the intensity used for each pixel of the real-space images, while the insets in the third column show the radial distributions.
The non-round `shadow' easily visible at the outer edges of the 1-bit diffraction pattern [Fig.~\ref{fig:bit_depth_atomic_resolution}(a)] is due to high angle cutoff in the microscope due to the image corrector.
Although the 1-bit diffraction patterns [Figs.~\ref{fig:bit_depth_atomic_resolution}(b) and \ref{fig:bit_depth_atomic_resolution}(c)] do not seem to contain much information, the ADF data [Fig.~\ref{fig:bit_depth_atomic_resolution}(a)] shows high quality atomic resolution imaging is possible, with the SrO, Ti, and O$_2$ columns \cite{Abramov_AC_1995_SrTiO3_structure} all resolved, as shown in the inset schematic.

The much higher frame rates achievable of 12,500 frames per second with 1-bit data in continuous read-write mode makes this acquisition configuration particularly suitable for navigation during setup or in especially beam sensitive materials.
One image of 256$\times$256 probe positions takes about 5 seconds to complete at this rate, but smaller scan sizes are often adequate for navigation.
In the experiment, the beam current was maintained and this unavoidably resulted in the central portion of the diffraction pattern being saturated when recording at bit depths of 1 and 6.
As shown by the radial distributions, we have selected the scattering angles where the detector is not saturated and contrast can be extracted.
With shorter exposures or lower beam currents, regions closer to the central spot of the diffraction pattern will not be saturated and would produce useable image contrast.
In 6-bit mode, more features of the diffraction pattern are visible than in the 1-bit mode and the darkfield image [Fig.~\ref{fig:bit_depth_atomic_resolution}(d)] is better defined.
This trend continues to the 12-bit mode [Fig.~\ref{fig:bit_depth_atomic_resolution}(g)] where the direct beam is no longer saturated, as shown in the inset to Fig.~\ref{fig:bit_depth_atomic_resolution}(i).
However, more atomic columns are present in the image as a result of larger spatial drift during the longer acquisition (9 or 10 Sr columns per row in the 12~bit data compared to 8 or 9 columns per row for the 1-bit data).
The principal benefit of higher bit depth imaging in this context is that a greater range of scattering angles may be used for virtual aperture imaging post acquisition and the signal-to-noise ratio (SNR) ratio is generally higher, even if there is a cost in acquisition time and consequent drift.

Selection of higher scattering angles by saturating the central spot in 1-bit and 6-bit modes is possible here because, unlike CCDs, the Medipix3 is not damaged by the very intense direct beam, and because the noise-free readout enables each single electron hit to be accurately recorded.
With very intense beams (approximately 1~MHz count rate per pixel), the electron arrival rate can exceed the counting rate of the detector; this does no harm to the detector, but electrons are missed and the counts no longer represent an accurate reflection of arrival rates (and even a little below this level, counting linearity is lost).
In this data, however, the beam current was not high enough to cause such an effect and the slight dips in intensity in the centre of Figs.~\ref{fig:bit_depth_atomic_resolution}(h) and \ref{fig:bit_depth_atomic_resolution}(i) are due to the real internal structure of the brightest portion of the diffraction pattern being resolved at the highest bit depth.

The fastest frame rate demonstrated above, corresponding to 80~$\muup$s per scan pixel, is still substantially slower than that of commonly used scintillator or photomultiplier tube based STEM detectors.
While switching between these detectors is a common procedure with little overhead, pixelated STEM detectors with higher speeds would be beneficial for the efficiency of live-imaging, and for low dose imaging of beam sensitive materials without having to reduce the beam current.
The limiting factor in the Medipix3 setup we used is the 120~MHz clock of the Merlin readout system \cite{Plackett_2013_merlin}.
The Medipix3 chip itself can be clocked to 200-250~MHz, potential allowing a doubling of acquisition speed.
Additionally, the Medipix3 chip allows readout of regions of interest (ROI), potentially allowing much higher sub-frame readout rates, but readout systems with this capability are not yet commonly available.

The pnCCD detector \cite{Ryll_2016_pnccd} is a radiation hard 264$\times$264 pixel CCD-based sensor with a full frame rate of 1000~fps.
The chip supports binning along one axis, allowing speeds of up to 4000~fps with 4$\times$ binning (264$\times$66 pixels).
However, it was recently reported that these speeds may be doubled with optimisation of the operation conditions and timing coordination of the readout ASICs \cite{huth_2019_MandM_pnccd_fps}.
By windowing to a 24 pixel wide strip, the detector is reported to operated at 10000~fps, which is approaching that of the 12500~fps 256$\times$256 1-bit data shown in this work.
At the maximum SNR mode of operation and imaging 200~keV electrons, the maximum number of primary electrons that may be measured per pixel is about 1 in the pnCCD detector.
This number rises to about 7 in the highest capacity mode of operation, but with a reduction in the SNR.
This detector does not benefit from the gapless or noise-free readout of the Medipix detector used here, but can still produce excellent results at low doses \cite{huth_2019_MandM_pnccd_fps}.

Alternative modes of operation can yield even faster data readouts from current generation detectors.
The Timepix3 chip \cite{Poikela_2014_JoI_Timepix3} operates with a 640~MHz clock, giving a timestamp resolution of 1.56~ns, and supports a data-driven mode of acquisition.
In this mode, only data from events are read out rather than the full array of pixel counts.
The data packet itself includes additional information such as time over threshold and time of arrival and, consequently, additional or alternative strategies are required to process this type of data (with the potential benefit of more advanced signal processing).
The maximum hit rate is 40~M hits/s/cm$^2$ ($\sim$80~M hits/s for a single 1.98~cm$^2$ chip) and provides faster readout in event driven mode than in frame mode for less than 50\% occupancy.
While this approach allows very short effective exposures, due to the increased size of the packet, the overall counting rate is reduced compared to Medipix3 detector.
However, unlike in the 1-bit Medipix3 data, the intensity distribution of the image signal would be accurately recorded without saturation in event driven modes of acquisition.

Beyond this, the collaboration behind the next-generation Medipix4 detector is targeting imaging rates which are compatible with human CT imaging \cite{Campbell_2016_mpx4}.
The typical detector dose varies in CT imaging, but can be of the order of 10$^3$ - 10$^4$ M hits/s/cm$^2$ \cite{Taguchi_2013_ct}, significantly higher than is possible in current generation Medipix detectors.

\begin{figure*}[hbt!]
  \centering
      \includegraphics[width=17cm]{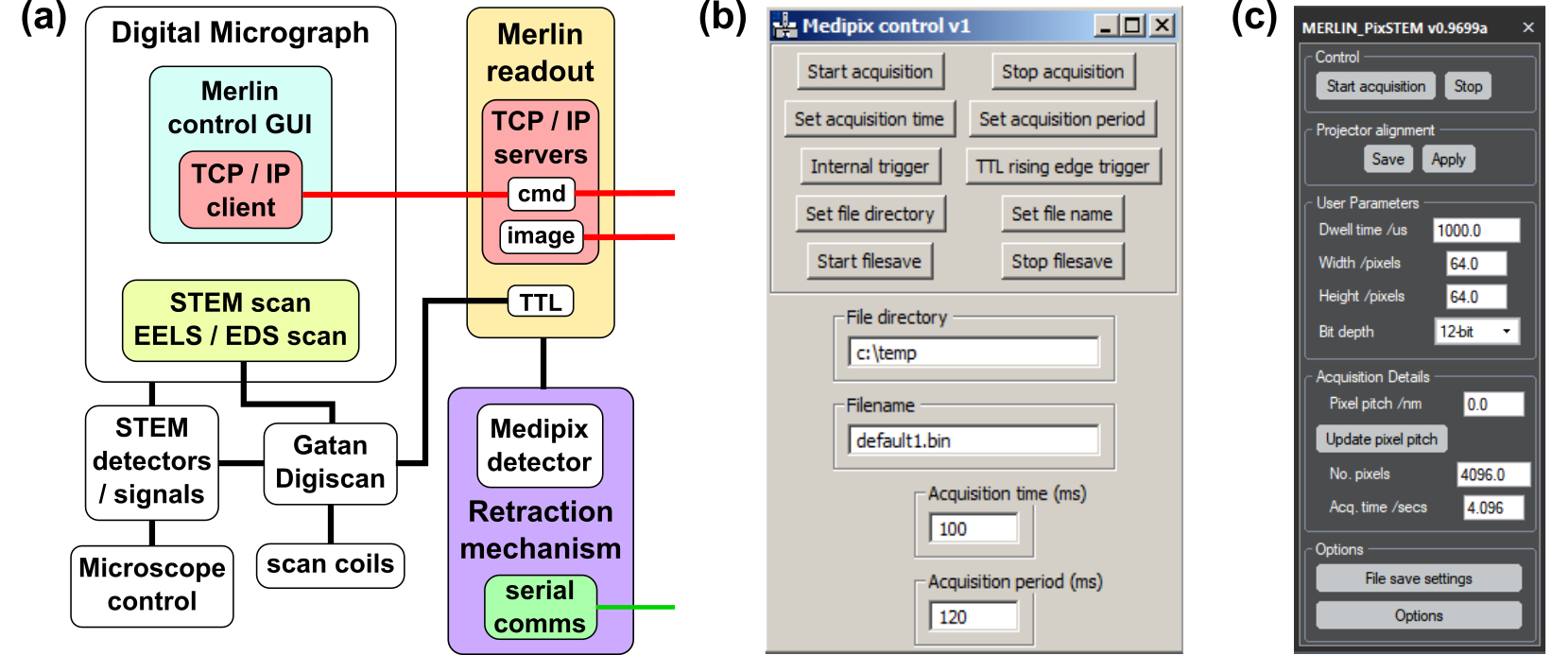}
      \caption{(a) Schematic of the Merlin Medipix3 readout control architecture and screenshots of (b) low level (\texttt{Merlin Control}) and (c) optimised continuous read-write mode (\texttt{MERLIN\_PixSTEM}) plugins for Gatan Digital Micrograph (DM). Note that the DM plugin in (c) is not related to the similarly named Python \texttt{pixStem} library.}
      \label{fig:hardware_live_visualisation_arch}
\end{figure*}

Regardless of the imaging mode used for the collection of the source data, smaller bit depths also makes both the file storage and the data processing more efficient; a 6-bit dataset is about 4 times smaller than a 24-bit one of the same scan area, making it much more convenient to store and transfer.
This advantage also extends to the data processing, since loading and processing data files which are 4 times smaller will be much quicker.

\section{Medipix3 Data Acquisition}
\label{sec:data_acquisition}
Data from the Medipix3 detector was acquired through the Merlin readout system \cite{Plackett_2013_merlin}.
This allows setting of the acquisition parameters, either through a graphical user interface (GUI) or over TCP/IP, and reads and processes the raw data through a field-programmable gate array (FPGA), returning the data to the acquisition computer.
The FPGA processing can be bypassed to some extent by operating the system in `raw' mode.
This enables larger scan sizes at high frame rates, with the requirement that the data must be reshaped post-acquisition, and with no live visualisation of the acquired images directly in the Merlin software.
However, the Merlin TCP/IP data API remains functional, so it is possible to get live imaging through other means (see Section~\ref{sec:live_processing}).

The Merlin system can be triggered by software over TCP/IP or by hardware (TTL) input.
We typically use the latter approach and couple to the TTL signals produced by a Gatan DigiScan system, as shown schematically in Fig.~\ref{fig:hardware_live_visualisation_arch}(a).
This produces extra acquisitions due to triggers sent during the flyback time and the handling of these is discussed in Section~\ref{sec:data_storage}.
The main advantage of this approach is that Gatan Digital Micrograph (DM), in addition to allowing access to microscope control, can be used for setting scan parameters in one of several ways discussed below, and additional STEM detector signals may be acquired simultaneously.

The simultaneously acquired DM datasets also serve to document the microscope and scan parameters in the data tags, which can then be used in data conversion (discussed in Section~\ref{sec:data_storage}), abstracting away the differences in how various microscope manufacturers provide microscope configuration information.

When regular STEM detectors can be used to navigate, the images produced from them may be used to set region of interest (ROI) scans using an image produced by a prior `survey' scan, following the spectrum imaging methodology, or regular STEM scans may be used to maximise read rates.
When these approaches are used, a scripted DM plugin may be used for setting low level Merlin parameters \cite{merlin_dm_plugin}, as shown in Fig.~\ref{fig:hardware_live_visualisation_arch}(b).
Alternatively, real space pixel sizes and scan ranges may be set and low level DM commands used to configure and enable the scan.
A scripted DM GUI has been developed, \texttt{MERLIN\_PixSTEM}, to coordinate this with configuring the Merlin system to acquire data in the optimised continuous read-write mode and is shown in Fig.~\ref{fig:hardware_live_visualisation_arch}(c).
Amongst the other features implemented, this plugin also allows different projection system settings to be saved and restored, enabling efficient switching between different detectors.
The Merlin system includes two TCP/IP servers, shown in red in Fig.~\ref{fig:hardware_live_visualisation_arch}(a), one for setting and reading acquisition parameters and the second for image data transfer.
The DM scripted GUIs, indicated in cyan in Fig.~\ref{fig:hardware_live_visualisation_arch}(a), interface with the Merlin communication server through a separate TCP/IP C++ plugin \cite{merlin_dm_plugin}, shown in red.
The TCP/IP plugin may be installed alone, allowing it to be used for many other communication purposes.
For more advanced control of the Merlin system over TCP/IP, a Python implementation of Merlin TCP/IP commands has been developed \cite{merlin_interface}.

\begin{figure}[hbt!]
  \centering
      \includegraphics[width=7.0cm]{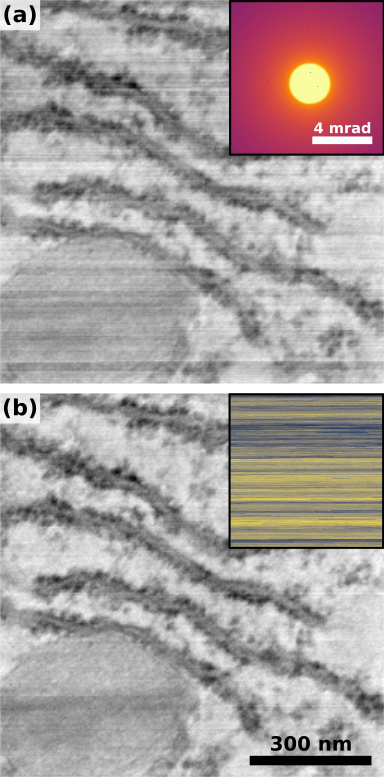}
      \caption{Gun noise correction in a bright field (BF) STEM image produced from a 4-D dataset from a thin fixed but unstained section of mouse liver, showing part of one cell including the end of a mitochondrion organelle and part of the rough endoplasmic reticulum.
      Measured (a) BF image and (b) the same image after gun noise correction using the \texttt{fpd.tem\_tools.nc\_correct} function.
      The inset to (a) shows the summed diffraction pattern on a logarithmic scale, while that to (b) shows the recorded gun noise.
      The acceleration voltage was 200~kV, the objective lens was off, the condenser aperture was 30~$\muup$m, the camera length was 600~cm, the convergence semi-angle was 13.1~mrad, and the pixel spacing was 3.7~nm.}
      \label{fig:nca_correction}
\end{figure}

An example of an additional STEM signal we collect is the STEM noise correction (NC) signal, which may be used for gun noise correction of the 4-D dataset.
The NC signal is produced by a current pickup attached to the condenser aperture and gives a measure of the gun emission.
A similar approach to gun signal measurement was recently reported and shown to have good linearity to the probe current \cite{house_schamp_yang_2018_ap_gun_noise}.
Correction of gun noise is particularly useful in intensity-based low contrast imaging modes, as shown in the bright field (BF) images of a mouse liver microtomed thin section in Fig.~\ref{fig:nca_correction}.
The image in Fig.~\ref{fig:nca_correction}(a) is produced by summing the entire diffraction pattern (an example is shown in the inset) at each scan position.
Sample contrast primarily arises due to incoherent Rutherford scattering of electrons to angles beyond the detector.
The large circular feature in the bottom left corner is part of a mitochondrion organelle, while the darker spotted stripe structures are endoplasmic reticula studded with ribosomes.

The horizontal stripes in the as-measured image in Fig.~\ref{fig:nca_correction}(a) are from short period variations in the cold-FEG emission.
The gun signal measured by the NC detector is shown in the inset to Fig.~\ref{fig:nca_correction}(b).
Figure~\ref{fig:nca_correction}(b) itself shows the corrected image produced by minimising the contrast introduced by these gun emission current variations using a linear gun-noise model.
This reveals much more detail in the BF image than previously seen in Fig.~\ref{fig:nca_correction}(a).
The residual horizontal features in the corrected image are most likely a result of variations in linearity of the NC detector signal from things like amplifier drift or external noise.
Taking the corrected image as a reference, the power SNR of the uncorrected image, calculated using the implementation of the two-image method \cite{Frank1980_2imageSNR} in the \texttt{fpd.utils} module, is 11~dB, giving a measure of the improvement in the image quality by applying gun noise correction.

The Medipix3 data acquired following the above methodologies can be saved to disk on the acquisition computer in a flat binary format and may also be sent over the network using TCP/IP.
Conversion of the binary data to more appropriate formats is discussed in Section~\ref{sec:data_storage}.
The network transfer of data has many potential uses and we discuss these in the context of live data processing in the next section.

\section{Live Data Processing}
\label{sec:live_processing}
Live feedback from the data collected by a fast pixelated detector in a STEM acquisition is crucial for both optimising imaging conditions and navigating to regions of interest in a sample.
This is especially true for some modes of imaging where traditional STEM detectors may not produce useful contrast, such as when imaging magnetic features which are typically not visible in STEM without a custom segmented detector and readout system \cite{mcgrouther2014development}.
To facilitate real-time feedback, we developed the Python library \texttt{fpd\_live\_imaging} \cite{fpd_live_imaging}, which implements multiple common analysis routines and wraps processing routines from other libraries \cite{fpd}.
Although the \texttt{fpd\_live\_imaging} package was developed for use with the Medipix3 detector and Merlin readout system, its design is modular and can easily be extended to work with any detector.
Our implementation takes advantage of the many cores available in modern CPUs by employing Python's multiprocessing library.
Shared parameters and data are passed between the separate processes through `queue' objects or other shared memory.
This approach enables good performance, even at very high data rates.

\begin{figure*}[hbt!]
  \centering
      \includegraphics[width=14.5cm]{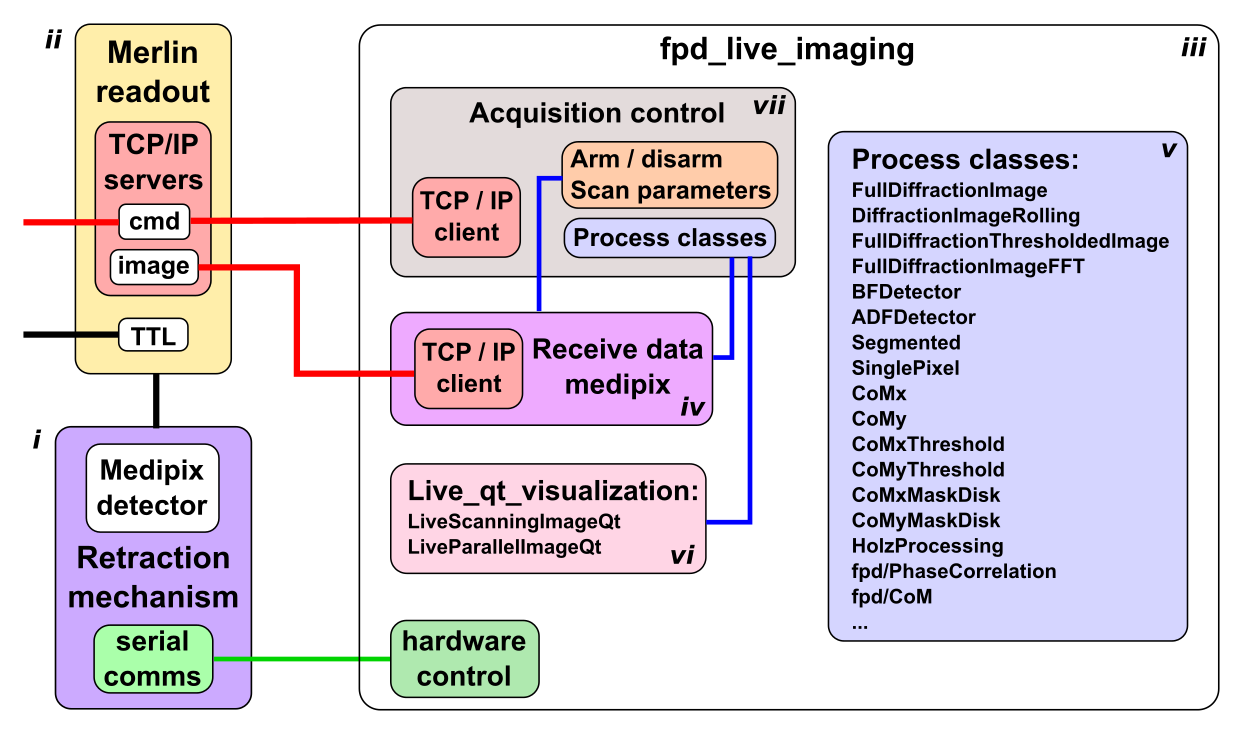}
      \caption{Schematic of the fast pixelated detector live visualisation library architecture, showing the relationship between the Medipix3 detector and retraction mechanism (purple, \textit{i}), the Merlin readout system (yellow, \textit{ii}) and the \texttt{fpd\_live\_imaging} library (white, \textit{iii}). Multiple processing classes (blue, \textit{v}) are implemented for scanning and imaging modes.}
      \label{fig:live_visualisation_arch}
\end{figure*}

The internal workings of the package are outlined in Fig.~\ref{fig:live_visualisation_arch}.
The Medipix3 1R insertion and retraction mechanism (shown coloured in purple, \textit{i}) is controllable through a serial interface (shown in green) and is made possible through library function calls.
As discussed in the previous section, the Merlin system (drawn in yellow, \textit{ii}) can be interfaced with via two TCP/IP servers (shown in red), which are utilised by the \texttt{fpd\_live\_imaging} package (white, \textit{iii}) to get data from the detector and to control the acquisition of data.
The first step in the visualisation is receiving the raw binary data from the Merlin TCP/IP data interface using the \texttt{receive\_data\_medipix} function (\textit{iv}).
This function runs a TCP/IP socket which gets the raw binary data and passes it along to a parser function.
The function has its own CPU process to be able to handle the very high framerate of the Medipix3 detector.
Due to the nature of the TCP/IP protocol, the raw binary images can be split into different fragments.
These fragments are pieced together in the parse function, which results in the image in the form of a NumPy array \cite{numpy}.
The function also handles the bit depth of the data and the number of pixels in the detector, and also runs in its own separate CPU process.

After having constructed the image in the form of the NumPy array, a copy is sent to any number of data processing classes.
These data processing classes are shown in blue (\textit{v}) in Fig.~\ref{fig:live_visualisation_arch} and can be separated into two categories based on the imaging mode: scanning and parallel.
The scanning data classes include things like virtual bright field and annular darkfield, where the input detector image is reduced to a single output value.
In the parallel data classes, the output image is the same size as the input one, and the processing methods include passing through the input image, a thresholded version of the input image, or a Fourier transformed image.
All these run in separate CPU processes.
In addition to the aforementioned processing classes are ones for single pixel extraction, centre of mass and phase-correlation for electro- or magneto-static field imaging, and routines for HOLZ analysis.
Virtual detector imaging and HOLZ data processing are covered in Part~II of this work \cite{methods_part2_arxiv}, while field mapping will be covered in Part~III.

The processing time varies greatly, depending on the computational complexity of the routine \cite{nord_magnetic_2016}, and the choice of routine depends upon the nature of the sample.
For example, in magnetic imaging, the integrated induction components perpendicular to the electron path can be determined from deflections in the position of the bright field disc \cite{CHAPMAN1999729}.
The centre of mass calculation provides good contrast in many cases but can be affected by the crystallinity of the sample due to intensity diffracted from the bright field disc to angles either outside or inside the detector collection angle \cite{chapman_1990_mdpc}.
Phase- or cross-correlation \cite{matus_pixelated_stem_magnetic_2016} approaches can greatly improve upon this at the expense of computation time, and can be crucial to detecting magnetic contrast in highly diffracting samples.
On the other hand, single pixel extraction, where a single pixel on the edge of the disc is used as a measure of up to around pixel-level disc shifts, requires the minimum of processing and is orders of magnitudes faster, taking approximately 2~$\muup$s when the 256$\times$256 scan position 12-bit dataset is in memory \cite{nord_magnetic_2016}.
As each selected pixel gives a measure of a component of the integrated induction in a direction tangential to the disc, the use of only two pixels out of each diffraction image is sufficient to form a qualitative 2-D vector map, which allows the user to at least navigate to an appropriate position, magnification and focus.
Multiple processes may be run sequentially or simultaneously, allowing the trade-off between runtime and sensitivity to be seen in real time.

\begin{figure}[hbt!]
  \centering
      \includegraphics[width=7.5cm]{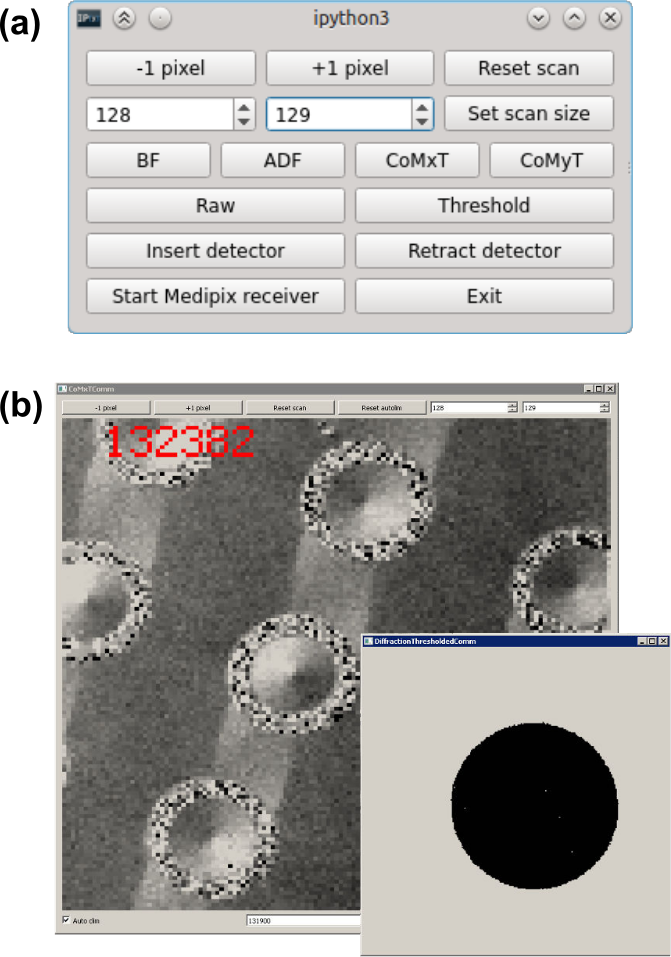}
      \caption{\texttt{fpd\_live\_imaging}'s graphical user interface, showing (a) the control window for the visualisation, and (b) a thresholded centre of mass contrast of a patterned 8~nm permalloy film capped with 4~nm of copper. The contrast in the 2~$\muup$m discs represents the beam deflection along a single axis, and shows that the discs support magnetic vortices. The inset to (b) shows the thresholded detector image.}
      \label{fig:live_visualisation_example}
\end{figure}

The output data from any kind of processing is sent to a visualisation class, which shows the result of the processing on the computer running the \texttt{fpd\_live\_imaging} package.
Due to rescaling of the intensity to optimise the contrast, this visualisation is qualitative, while the calculations themselves can be quantitative.
This computer may be anywhere on the network.
The visualisation is separated into parallel and scanning modes, as shown in pink (\textit{vi}) in Fig.~\ref{fig:live_visualisation_arch}, and they also run in separate CPU processes.
An example of the visualisation GUI is shown in Fig.~\ref{fig:live_visualisation_example}(b).
In this case, the image is from thresholded centre of mass analysis of data from a patterned DC sputtered 8~nm permalloy film capped with 4~nm of copper.
The 2~$\muup$m discs were patterned with a Ga focused ion beam, and the contrast in the resulting structures shows they support magnetic vortices.
A detailed study of the sample will be published elsewhere.
The GUI has buttons for setting the brightness and contrast during the acquisition, and the analysis parameters can be tuned during imaging, allowing for live optimisation of the required contrast.
Alternatively, the processed data can be sent over TCP/IP to any computer on the network, for example, directly into Digital Micrograph.

All the above processes are orchestrated from the `Acquisition Control' class (shown in brown, \textit{vii} in Fig.~\ref{fig:live_visualisation_arch}), which handles the initialisation and connection of all of these separate functions.
For ease-of-use, the Acquisition Control class can be accessed through a GUI, as shown in Fig.~\ref{fig:live_visualisation_example}(a).
This allows for starting and stopping of the acquisition, modification of the scan parameters, the addition and removal of processing classes, modification of their parameters, and insertion and retraction of the detector itself.

The three separate stages described above, reading data from the detector, processing the images, and visualising or sending the result over TCP/IP, are implemented in modular design, making simple the addition of new detector data sources, image processing classes, and visualisation.

\section{Data Storage}
\label{sec:data_storage}
The principal issues when choosing a file format for fast pixelated detector data are common across data from all detectors: the ability to store the data with the dimensionality of the scan, store metadata along with the detector data, allow access to subsets of the data without reading the entire and often very large dataset into memory, support compression, and be an open format with read and write support across a variety of programming languages.
An HDF5 \cite{hdf5_file_format} based format was chosen for our use since it meets all of the above requirements.

The HDF format has long been widely used in the synchrotron community and is increasingly being used in electron microscopy \cite{hyperspy_library, emd_format, pycroscopy}.
It can be both read and written in a number of programming languages, including MatLab, C++, Python, Java, R, and Gatan Digital Micrograph through a third party plugin \cite{DM_hdf5}.  
The HDF5 format consists of an arbitrary structure of hierarchies of groups containing further groups or datasets, enabling the relationship between data to be indicated by the file structure.
For datasets, the data type definitions are stored with the data, making it self-describing and ensuring maximum portability.
Additionally, all groups and datasets can have attributes, allowing user and acquisition metadata to be stored along with the detector data in appropriate locations.
The datasets may be of any number of dimensions and so it is ideal for multidimensional data from fast pixelated detectors when used in STEM or other acquisition modes.

\begin{figure}[hbt!]
  \centering
      \includegraphics[width=7.5cm]{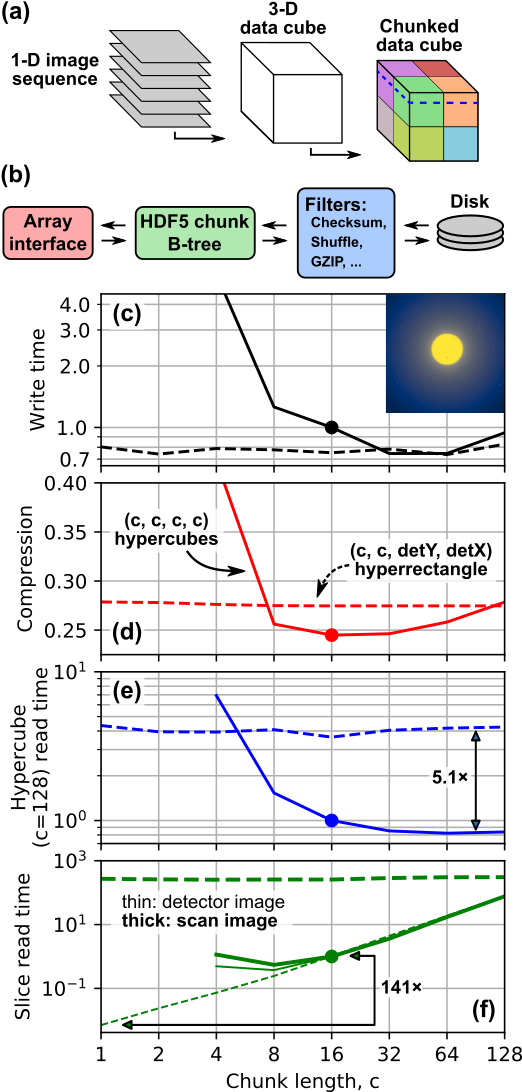}
      \caption{(a) Example of potential dataset chunking for data from a 1-D scan stored in an HDF5 file. (b) Data indexing sequence for chunked data. (c)-(f) HDF5 chunk performance metrics for the 256$\times$256 probe position STEM dataset from Fig.~\ref{fig:nca_correction} with a 256$\times$256 Medipix3 detector in 12-bit mode. Level 4 GZIP compression was used. All panels show metrics for hypercube chunks (solid lines) and hyperrectangle chunks (dashed ines) with dimensions matching those of the detector (detY, detX). The inset to (c) shows a diffraction image on a logarithmic scale. The (e, f) read and (c) write times are the ratios of the values to those for the hypercube chunk length of 16 (marked by symbols). The compression ratios (d) are of the entire HDF5 file relative to only the raw Merlin binary file. The read times in (e) and (f) are those required to load a 128-sided hypercube or single slices of the dataset, respectively, into an in-memory NumPy array using h5py.}
      \label{fig:chunks_chunk_performance}
\end{figure}

HDF5 has in-built support for a variety of compression algorithms and other so-called `filters', all providing transparent read and write access to the data.
To allow access to subsets of the data without having to decompress the entire dataset, the dataset can be divided into smaller pieces and stored in a B-tree, a balanced hierarchical data structure, by enabling `chunking'.
Figure~\ref{fig:chunks_chunk_performance}(a) shows an example of the potential chunking of a one-dimensional (1-D) scan dataset.
The stack of images (shown on the left) occupy a 3-D data `cube' (middle), with one axis being the scan dimension.
On the right of panel (a) we show the same dataset with two chunks along each dimension, with each chunk in a different colour.
The dataset access sequence is summarised in Fig.~\ref{fig:chunks_chunk_performance}(b).
When indexing a chunked dataset, the B-tree is navigated and each chunk containing the required data is decompressed and only the selected components are returned.
For example, when reading the image slice shown by the blue dashed line in the right of Fig.~\ref{fig:chunks_chunk_performance}(a), each of the top four chunks must be read.

\begin{figure*}[hbt!]
  \centering
      \includegraphics[width=14cm]{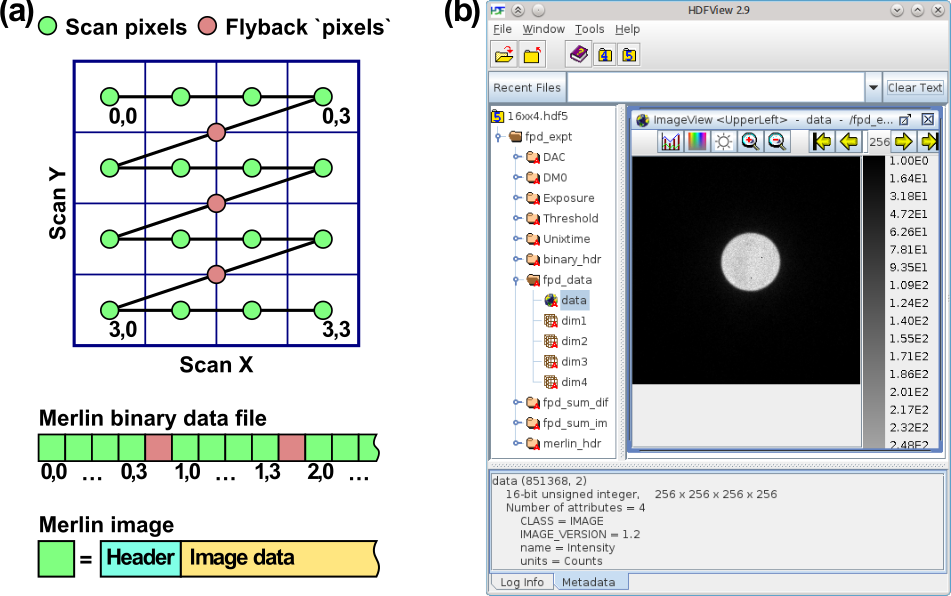}
      \caption{(a) Example of additional `flyback' pixels in a 2-D STEM scan, and the on-disk Merlin binary file structure. (b) Overview of our HDF5 file structure, shown in HDFView \cite{HDFView}, using the same data as used for Fig.~\ref{fig:nca_correction}.}
      \label{fig:fly_file}
\end{figure*}

\subsection{Chunk Size}
When choosing a chunk size, a compromise is made between the cost of B-tree navigation, compression level, and data reading speed, with the optimum choice ultimately depending on the intended data access pattern.
For STEM data, the diffraction pattern can be sparse and compression can be optimised by chunking in both the scan and image dimensions.
Although most of the issues discussed in this section will be common to all detectors, we note that the Medipix detector has three relevant features which separate it from most others.
The first two are that the detector can unambiguously detect single electron impacts, and that the data is read out free of addition noise, meaning that dark regions of images can be truly filled with zeros rather than noise.
The third point is that data from the Medipix3 detector is zero padded to align it with common data types (e.g. 12-bit data is stored as 16-bit), allowing compression to achieve significant reductions in file size.
For example, with a chunking of 16 along each axis of a 4-D dataset, the in-built lossless GZIP compression at level 4 typically reduces the data size of a scanning acquisition of 256$\times$256 probe positions in 12-bit mode from 8.6~GB to 2.7~GB.

Many chunking strategies are possible and, here, we explore two of them: a hypercube with equal chunk sizes across all axes; and a hyperrectangle with equal chunk sizes along the scan axes, and detector axes chunk sizes matching the detector dimensions.
The first approach gives the most uniform data access properties across different axes and, importantly, can improve data processing efficiency by allowing reduction of the volumes of data that must be read for some analyses.
For example, with a chunk size of 16 along each axis, getting the direct beam in a dataset where it resides in four chunks would require loading into memory only those chunks, corresponding to only 1.6\% of the total file.
The second approach of hyperrectangles is more suited to applications which only ever access full images, since requesting even a single pixel from an image would require the entire image to be read by the HDF5 library and, similarly, indexing a 2-D slice perpendicular to the detector axis would require the entire dataset to be read, resulting in significant overheads.

Figures~\ref{fig:chunks_chunk_performance}(c)-\ref{fig:chunks_chunk_performance}(f) show three HDF5 performance metrics as a function of the chunk edge length for the two chunking approaches (hypercube: solid lines; hyperrectangle: dashed lines) using the liver sample data in Fig.~\ref{fig:nca_correction}.
The three metrics are normalised write time [Fig.~\ref{fig:chunks_chunk_performance}(c)], compression ratio [Fig.~\ref{fig:chunks_chunk_performance}(d)] and normalised read time [Fig.~\ref{fig:chunks_chunk_performance}(e, f)].
The write and read times were normalised with respect to the values for the hypercube chunk length of 16 (marked by symbols), while the compression ratio is with respect to the size of the raw Merlin data file.
Hypercube chunks of length 16 is used as the reference because it is the default value in our implementation, as a result of it being a reasonable compromise for most applications.

For hyperrectangle chunks, the write time [Fig.~\ref{fig:chunks_chunk_performance}(c)], compression ratio [Fig.~\ref{fig:chunks_chunk_performance}(d)] and read time for a 128 side length hypercube [Fig.~\ref{fig:chunks_chunk_performance}(e)] are mostly independent of chunk size, whereas there are significant variations for hypercubes.
As the chunk size is increased, all three metrics start very high (poor) due to the very large number of chunks, and then either go through a minimum at a chunk length around 16-32 (write time and compression) before increasing slightly, or plateau (hypercube read time).
With all else being equal, the hypercube read time for the largest chunk sizes in Fig.~\ref{fig:chunks_chunk_performance}(e) should be 4$\times$ worse for the hyperrectangle than for the hypercube, due to the overhead from the HDF5 library having to read the entire dataset, and the actual value of 5.1 is close to this.
Similarly, reading a subset of any of the chunk sizes shown in Fig.~\ref{fig:chunks_chunk_performance}(e) will result in performance reduction, with smaller chunk sizes being less affected.

In Fig.~\ref{fig:chunks_chunk_performance}(f) we show the read times for indexing a single slice of the dataset, with the slice creating a detector image (thick lines) or a scan image (thin lines).
As explained above, when indexing across a non-detector axis, the entire data set must be read and this is the reason for the high and chunk size independent read times for the hyperrectangle approach (thick dashed line).
Above a chunk length of around 16, all other datasets lie on top of one another and follow a linear relationship.
Below this point, the hypercube read times begin to plateau and then increase, while the hyperrectangle chunked read times maintain the linear trend all the way to chunk lengths of 1.
At this chunk size, reading a single image is 141$\times$ faster than for our default setting of length 16 hyperchunks, which is not as fast as the 256$\times$ smaller data would predict, due to additional overheads.
Thus, for reading single images, having chunks of single images is a simple strategy to maximise performance, at the expense of flexibility in reading the data in other ways.
However, a very simple method that often allows for similar read speed while maintaining flexibility, is to read images (when they are needed as full images) from a hypercube chunked dataset in numbers that are aligned to a chunk size.
This is the approach of data processing using the \texttt{fpd} package, which is discussed in more detain in Part~II \cite{methods_part2_arxiv}.
For the example in Fig.~\ref{fig:chunks_chunk_performance}(f), taking length 16 hypercubes as the reference, reading 16$\times$16 images with hyperrectangles of length 1 takes 0.95$\times$ the reference time, whilst reading the same with hyperrectangles of length 16 takes 1.12$\times$ the reference time.
For markedly different datasets or where the data access pattern is known in advance, the optimum chunking may be somewhat different and this can be set by the user at the point of conversion.

\subsection{Merlin Data}
The Merlin readout software stores the detector and readout system parameters in a separate header file, and the detector data as a stream of uncompressed binary data, with each image containing a variable length header of acquisition parameters specific to that image.
The \verb|MerlinBinary| class from the \verb|fpd_file| module of the \verb|fpd| library \cite{fpd, fpd_demos} allows parsing of data files, array access to raw data using memory mapping, and conversion to the HDF5 format.
The scan parameters and metadata can be extracted from Digital Micrograph files acquired simultaneously with the diffraction patterns, or may be supplied separately.
For the former case, the DM files are accessed through the HyperSpy library \cite{hyperspy_library} and are also embedded in the HDF5 file as raw binary blobs for reuse in the proprietary DM software.
All DM files are also stored in the HDF5 file in the open EMD format \cite{emd_format} (discussed in the next section).
Examples of Merlin data converted to HDF5 format using the \verb|MerlinBinary| class are available in the open data deposit for this work \cite{paper_dataset_zenodo}.

Figure~\ref{fig:fly_file}(a) shows an example of a 2-D scan when the acquisition is being triggered by the microscope scanning system.
Images, indicated in green, are acquired on a regular scan grid.
As discussed in Section~\ref{sec:data_acquisition}, during the time when the beam is being moved from the end of one row to the start of the next, the `flyback' time, the DigiScan system continues to send triggers, causing additional images to be acquired.
These are shown in red, and may be excluded during data access and conversion to the HDF5 format with appropriate parameter settings.

The image data in the Merlin binary file is in C-order, that is, with the fastest moving index being in the last dimension, as depicted in Figure~\ref{fig:fly_file}(a).
The HDF5 library is a self-describing one and returns datasets stored within it in the form appropriate for the library being used, but stores the data internally in C-order.
C-order is also the default ordering in NumPy and, thus, we naturally store the multi-dimensional pixelated STEM datasets in C-order, with the first axes being the scan ones and the last two being the detector ones.
Most pixelated STEM datasets are 4-D and of the type described.
However, the Medipix3 detector can be operated in colour mode, where an additional axis representing multiple thresholds exists between the scan and detector axes.
This axis, while not generally used in STEM acquisitions at present, is used for spectroscopic X-ray imaging, is useful for characterising the detector performance using X-rays, and is supported by the \verb|fpd| library.

While the HDF5 conversion is most appropriate for data archival and later processing of data acquired under all modes of operation, the \verb|MerlinBinary| class also provides a memory mapped array interface to the data on disk for most but not all acquisitions.
For example, 1-bit data acquired in raw mode is stored as 1-bit by the Merlin system and has the image segments out of order, and cannot currently be easily memory mapped.
However, in most cases, this mode of access allows the dataset to be visualised or processed without conversion of the data on disk to the HDF5 format, which is particularly useful for checking datasets immediately after acquisition.

\subsection{HDF5 File Structure}
Figure~\ref{fig:fly_file}(b) shows an overview of one of the HDF5 files read in HDFView \cite{HDFView}, a Java GUI program that allows, amongst many other things, quick inspection of HDF file contents.
Information from the binary headers such as the DAC values, the image exposure, the comparator threshold values and the acquisition time are automatically extracted and included as datasets in the HDF5 file.
During conversion to the HDF5 format, a sum of both the image and diffraction dimensions are generated and stored in the HDF5 file as separate datasets, resulting in bright field and diffraction sum images.
These images may be used for data inspection and navigation without having to process the entire dataset in order to re-render them every time the file is loaded.
Also during conversion, any bad pixels in the detector (hot, noisy or dead), may be replaced by interpolated values when the user supplies a mask image, which can be important for optimisation of some forms of data analysis.
Additionally, an arbitrary function may be applied to each image during conversion.
This has many possible uses, including to correct for unwanted image shifts due to descan and to apply image transforms to correct image distortion.

The detector data, including images created from it, is stored in the EMD format, a simple open subset of the HDF5 format, created by a specific collection of datasets and attributes \cite{emd_format}.
The EMD datasets may be read in software such as EMD viewer \cite{emd_viewer}, HyperSpy \cite{hyperspy_library} and, of course, any HDF5 reader.
Many utility functions are also provided in the \texttt{fpd.fpd\_file} module, allowing conversion to other formats and extraction of data.
These include ones to access datasets as a namedtuple (\texttt{fpd\_to\_tuple}) which allows for indexing or tab completion, and one for accessing the same as HyperSpy objects (\texttt{fpd\_to\_hyperspy}).

Other projects have also adopted or are compatible with the EMD format \cite{hyperspy_library, py4dstem}, and we note that work is currently underway by the LiberTEM project \cite{nexus_libertem} to include (transmission) electron microscopy data in the NeXus data format \cite{nexus_format}.
The Nexus format is another open subset of the HDF5 file format, originally developed to improve the data exchange within the fields of neutron, X-ray and muon experiments.
Having a common data format across all of these fields would be beneficial, since it would make the sharing of data and of processing routines that rely on metadata easier than is currently the case.
This format could clearly be used within the suite of tools described in this paper.

\subsection{Merlin Equipped SPED Systems}
A recent development in precession electron diffraction \cite{Midgley_IUCrJ_2015_ped_rev} (PED) is the use of fast pixelated detectors.
One such example is the work by NanoMEGAS to incorporate a Medipix3 DED into their DigiSTAR precession system in order to enable high fidelity recording of diffraction patterns in scanning PED (SPED) applications, as has been tested in recent work \cite{MacLaren_MandM_2020_sped}.
Additional benefits brought about by the use of DEDs in SPED will be discussed in Part~II of this work \cite{methods_part2_arxiv}.
We note here, however, that the properties of the 4-D dataset obtained by such a system are in many ways equivalent to those of 4-D non-SPED datasets, and so many of the same issues of data access, storage, and processing apply here too.
To enable these datasets to be more easily used, the \texttt{topspin\_app5\_to\_hdf5} function of the \texttt{fpd.fpd\_io} module allows conversion of data originally recorded in the native NanoMEGAS TopSpin app5 format to one almost identical to the HDF5 format outlined above.
Alternatively, the Merlin acquisition software can be programmed to output the data directly to a raw file whilst acquisition is being performed and controlled by the TopSpin software.
The main differences between the converted files is the inclusion of precession metadata instead of Medipix3 metadata in SPED datasets, and the absence of simultaneously acquired DM datasets.

\section{Summary}
The use of fast pixelated detectors for electron imaging is a burgeoning field with the prospect of revolutionising many aspects of transmission electron microscopy (TEM) and, in particular, scanning TEM.
We have presented many of the key tools needed to i) acquire data from fast pixelated detectors, ii) analyse in real-time the data from one and visualise the results, and iii) store data from them in an optimised way.

The software packages presented are hosted in public repositories \cite{merlin_dm_plugin, merlin_interface, fpd_live_imaging, fpd, fpd_demos, pixstem}, are under active development and contain many more features than are covered in this short publication.
Many of the data analysis algorithms in these libraries are applicable to data from any detector.
Most of these packages are provided under an open source licence, allowing transparency of the algorithms implemented and for them to be continually improved upon by the community.

Part~II of this paper will cover post-acquisition processing and visualisation of data from fast pixelated detectors, with examples of their application to the study of the structure of materials studied using scanning transmission electron microscopy \cite{methods_part2_arxiv}.
A final part~III will cover aspects related to differential phase contrast analysis.

\section*{Acknowledgements and author contributions}
G.W.P.and M.N. were the principal authors of the libraries reported herein (details of all contributions are documented in the repositories), and have made almost all of these available under open source licence GPLv3 for the benefit of the community. R.W.H.W. and K.A.P. have also made contributions to the source codes in these libraries.
G.W.P and M.N. have led the drafting of this manuscript.
The performance of this work was mainly supported by Engineering and Physical Sciences Research Council (EPSRC) of the UK via the project ``Fast Pixel Detectors: a paradigm shift in STEM imaging'' (grant No. EP/M009963/1). 
G.W.P. received additional support from the EPSRC under grant No. EP/M024423/1. 
M.N. received additional support for this work from the European Union’s Horizon 2020 research and innovation programme under the Marie Skłodowska-Curie grant agreement No 838001.

R.W.H.W., S.McV., I.M., K.A.P.and D.McG. have all contributed through acquisition and analysis of data and through participation in the revision of the manuscript.
The studentship of R.W.H.W. was supported by the EPSRC Doctoral Training Partnership grant No. EP/N509668/1. 
S.McV. was supported by EPSRC grant No. EP/M024423/1. 
I.M. was supported by EPSRC grant No. EP/M009963/1. 
The studentship of K.A.P. was funded entirely by the UK Science and Technology Facilities Council (STFC) Industrial CASE studentship ``Next2 TEM Detection'' (No. ST/P002471/1) with Quantum Detectors Ltd. as the industrial partner.
D.McG. was also supported by EPSRC grant No. EP/M009963/1. 
As an inventor of intellectual property related to the MERLIN detector hardware, he is a beneficiary of the license agreement between University of Glasgow and Quantum Detectors Ltd.

We thank Diamond Light Source, especially Dr. Chris Allen and Prof. Angus Kirkland, for access and support in the use of the electron Physical Science Imaging Centre E02 microscope (proposal number EM16952) that provided the atomic resolution SrTiO3 data presented here;
Quantum Detectors Ltd. for Medipix3 detector support; 
Dr. Bruno Humbel from Okinawa Institute of Science and Technology and Dr. Caroline Kizilyaprak from the University of Lausanne for providing the liver sample; 
Dr. Kayla Fallon from the University of Glasgow for the sample data shown in Fig. 5(b), and Dr. Sinan Azzawi and Prof. Del Atkinson from University of Durham for providing the sample;
Dr. Ingrid Hallsteinsen and Prof. Thomas Tybell from the Norwegian University of Science and Technology (NTNU) for providing the SrTiO$_3$ sample shown in Fig. 1;
and NanoMEGAS for the loan of the DigiSTAR precession system and TopSpin acquisition software.
Development of the integration of TopSpin with the Merlin readout of the Medipix3 camera has been performed with the aid of financial assistance from the EPSRC under grant No. EP/R511705/1 and through direct collaboration between NanoMEGAS and Quantum Detectors Ltd.

\section*{Data and Software Availability}
The source data and scripts to analyse the data and produce the results presented here are publicly available \cite{paper_dataset_zenodo}.
The software presented in this work is available in public Git repositories \cite{merlin_dm_plugin, merlin_interface, fpd_live_imaging, fpd, fpd_demos, pixstem}, which also document the source of contributions to the code.
Enquiries about the \texttt{MERLIN\_PixSTEM} DM plugin should be directed to \texttt{Damien.McGrouther@glasgow.ac.uk}.

\bibliographystyle{mm_nat}

\begin{thebibliography}{88}
\providecommand{\natexlab}[1]{#1}
\providecommand{\url}[1]{\texttt{#1}}
\expandafter\ifx\csname urlstyle\endcsname\relax
  \providecommand{\doi}[1]{doi: #1}\else
  \providecommand{\doi}{doi: \begingroup \urlstyle{rm}\Url}\fi

\bibitem[Abramov et~al.(1995)Abramov, Tsirelson, Zavodnik, Ivanov, and
  Brown]{Abramov_AC_1995_SrTiO3_structure}
Abramov, Y.~A., Tsirelson, V.~G., Zavodnik, V.~E., Ivanov, S.~A., and Brown,
  I.~D., (1995).
\newblock {The chemical bond and atomic displacements in SrTiO$_3$ from X-ray
  diffraction analysis}.
\newblock \emph{Acta Cryst. B}, {\bfseries {\bfseries 51}\penalty0 (6)},
  \penalty0 942--951.
\newblock \doi{10.1107/S0108768195003752}.

\bibitem[Ballabriga et~al.(2013)Ballabriga, Alozy, Blaj, Campbell, Fiederle,
  Frojdh, Heijne, Llopart, Pichotka, Procz, Tlustos, and
  Wong]{Ballabriga_2013_Medipix3RX}
Ballabriga, R., Alozy, J., Blaj, G., Campbell, M., Fiederle, M., Frojdh, E.,
  Heijne, E. H.~M., Llopart, X., Pichotka, M., Procz, S., Tlustos, L., and
  Wong, W., (feb 2013).
\newblock {The Medipix3RX: a high resolution, zero dead-time pixel detector
  readout chip allowing spectroscopic imaging}.
\newblock \emph{J. Instrum.}, {\bfseries {\bfseries 8}\penalty0 (02)},
  \penalty0 C02016--C02016.
\newblock \doi{10.1088/1748-0221/8/02/c02016}.

\bibitem[Banerjee et~al.(2017)Banerjee, Baker, Doye, Nord, Heath, Erotokritou,
  Bosworth, Barber, MacLaren, and Hadfield]{banerjee_fem_2017}
Banerjee, A., Baker, L.~J., Doye, A., Nord, M., Heath, R.~M., Erotokritou, K.,
  Bosworth, D., Barber, Z.~H., MacLaren, I., and Hadfield, R.~H., (2017).
\newblock Characterisation of amorphous molybdenum silicide ({MoSi})
  superconducting thin films and nanowires.
\newblock \emph{Supercond. Sci. Technol.}, {\bfseries {\bfseries 30}\penalty0
  (8)}, \penalty0 084010.
\newblock \doi{10.1088/1361-6668/aa76d8}.

\bibitem[Behnel et~al.(2011)Behnel, Bradshaw, Citro, Dalcin, Seljebotn, and
  Smith]{cython}
Behnel, S., Bradshaw, R., Citro, C., Dalcin, L., Seljebotn, D.~S., and Smith,
  K., (2011).
\newblock Cython: The best of both worlds.
\newblock \emph{Comput. Sci. Eng.}, {\bfseries {\bfseries 13}\penalty0 (2)},
  \penalty0 31--39.
\newblock \doi{10.1109/MCSE.2010.118}.

\bibitem[Campbell et~al.(2016)Campbell, Alozy, Ballabriga, Frojdh, Heijne,
  Llopart, Poikela, Tlustos, Valerio, and Wong]{Campbell_2016_mpx4}
Campbell, M., Alozy, J., Ballabriga, R., Frojdh, E., Heijne, E., Llopart, X.,
  Poikela, T., Tlustos, L., Valerio, P., and Wong, W., (2016).
\newblock Towards a new generation of pixel detector readout chips.
\newblock \emph{J. Instrum.}, {\bfseries {\bfseries 11}\penalty0 (01)},
  \penalty0 C01007.
\newblock \doi{10.1088/1748-0221/11/01/c01007}.

\bibitem[Chapman and Scheinfein(1999)]{CHAPMAN1999729}
Chapman, J.~N. and Scheinfein, M.~R., (1999).
\newblock Transmission electron microscopies of magnetic microstructures.
\newblock \emph{J. Magn. Magn. Mater.}, {\bfseries {\bfseries 200}\penalty0
  (1)}, \penalty0 729 -- 740.
\newblock \doi{10.1016/S0304-8853(99)00317-0}.

\bibitem[Chapman et~al.(1990)Chapman, McFadyen, and McVitie]{chapman_1990_mdpc}
Chapman, J.~N., McFadyen, I.~R., and McVitie, S., (1990).
\newblock Modified differential phase contrast lorentz microscopy for improved
  imaging of magnetic structures.
\newblock \emph{IEEE Trans. Magn.}, {\bfseries {\bfseries 26}\penalty0 (5)},
  \penalty0 1506--1511.
\newblock \doi{10.1109/20.104427}.

\bibitem[Chapman et~al.(1978)Chapman, Batson, Waddell, and
  Ferrier]{Chapman1978_dpc}
Chapman, J., Batson, P., Waddell, E., and Ferrier, R., (1978).
\newblock The direct determination of magnetic domain wall profiles by
  differential phase contrast electron microscopy.
\newblock \emph{Ultramicroscopy}, {\bfseries 3}, \penalty0 203 -- 214.
\newblock \doi{10.1016/S0304-3991(78)80027-8}.

\bibitem[Clausen et~al.(2019)Clausen, Weber, @probonopd, Caron, Nord,
  Müller-Caspary, Ophus, Dunin-Borkowski, Ruzaeva, Chandra, Shin, and van
  Schyndel]{libertem}
Clausen, A., Weber, D., @probonopd, Caron, J., Nord, M., Müller-Caspary, K.,
  Ophus, C., Dunin-Borkowski, R., Ruzaeva, K., Chandra, R., Shin, J., and van
  Schyndel, J., (October 2019).
\newblock Libertem/libertem: 0.2.2.
\newblock \url{https://doi.org/10.5281/zenodo.3489385}.

\bibitem[Clough et~al.(2014)Clough, Moldovan, and Kirkland]{DirectDetectorsEM}
Clough, R.~N., Moldovan, G., and Kirkland, A.~I., (2014).
\newblock Direct detectors for electron microscopy.
\newblock \emph{J. Phys. Conf. Ser. I}, {\bfseries 522}, \penalty0 012046.
\newblock \doi{10.1088/1742-6596/522/1/012046}.

\bibitem[Crewe(1970{\natexlab{a}})]{Crewe1970_highResBio}
Crewe, A.~V., (1970{\natexlab{a}}).
\newblock High resolution scanning microscopy of biological specimens.
\newblock \emph{Ber. Bunsenges. Phys. Chem.}, {\bfseries {\bfseries
  74}\penalty0 (11)}, \penalty0 1181--1187.
\newblock \doi{10.1002/bbpc.19700741117}.

\bibitem[Crewe(1970{\natexlab{b}})]{crewe_1970_QRevBiophys}
Crewe, A.~V., (1970{\natexlab{b}}).
\newblock The current state of high resolution scanning electron microscopy.
\newblock \emph{Q. Rev. Biophys.}, {\bfseries {\bfseries 3}\penalty0 (1)},
  \penalty0 137–175.
\newblock \doi{10.1017/S0033583500004431}.

\bibitem[Crewe et~al.(1968)Crewe, Wall, and Welter]{Crewe_1968_JAP_stem}
Crewe, A.~V., Wall, J., and Welter, L.~M., (1968).
\newblock A high--resolution scanning transmission electron microscope.
\newblock \emph{J. Appl. Phys.}, {\bfseries {\bfseries 39}\penalty0 (13)},
  \penalty0 5861--5868.
\newblock \doi{10.1063/1.1656079}.

\bibitem[Crewe(1966)]{Crewe1966_science}
Crewe, A.~V., (1966).
\newblock Scanning electron microscopes: {Is} high resolution possible?
\newblock \emph{Science}, {\bfseries {\bfseries 154}\penalty0 (3750)},
  \penalty0 729--738.
\newblock \doi{10.1126/science.154.3750.729}.

\bibitem[de~la Pe{\~n}a et~al.(2018)de~la Pe{\~n}a, Ostasevicius, Fauske,
  Burdet, Prestat, Jokubauskas, Nord, Sarahan, MacArthur, Johnstone, Taillon,
  Caron, Migunov, Furnival, Eljarrat, Mazzucco, Aarholt, Walls, Slater,
  Winkler, Martineau, Donval, McLeod, Hoglund, Alxneit, Hjorth, Henninen,
  Zagonel, Garmannslund, and 5ht2]{hyperspy_library}
de~la Pe{\~n}a, F., Ostasevicius, T., Fauske, V.~T., Burdet, P., Prestat, E.,
  Jokubauskas, P., Nord, M., Sarahan, M., MacArthur, K.~E., Johnstone, D.~N.,
  Taillon, J., Caron, J., Migunov, V., Furnival, T., Eljarrat, A., Mazzucco,
  S., Aarholt, T., Walls, M., Slater, T., Winkler, F., Martineau, B., Donval,
  G., McLeod, R., Hoglund, E.~R., Alxneit, I., Hjorth, I., Henninen, T.,
  Zagonel, L.~F., Garmannslund, A., and 5ht2, (2018).
\newblock hyperspy/hyperspy: {HyperSpy} 1.3.1.
\newblock \url{https://doi.org/10.5281/zenodo.1221347}.

\bibitem[Dekkers and de~Lang(1977)]{Dekkers1977_dpc}
Dekkers, N. and de~Lang, H., (1977).
\newblock A detection method for producing phase and amplitude images
  simultaneously in a scanning transmission electron microscope.
\newblock \emph{Philips Tech. Rev.}, {\bfseries {\bfseries 37}\penalty0 (1)},
  \penalty0 1 -- 9.

\bibitem[Delpierre(2014)]{Delpierre_2014_hybrid_pix_dev}
Delpierre, P., (2014).
\newblock A history of hybrid pixel detectors, from high energy physics to
  medical imaging.
\newblock \emph{J. Instrum.}, {\bfseries {\bfseries 9}\penalty0 (05)},
  \penalty0 C05059.
\newblock \doi{10.1088/1748-0221/9/05/c05059}.

\bibitem[Donald and Craven(1979)]{Donald1979_pma}
Donald, A.~M. and Craven, A.~J., (1979).
\newblock A study of grain boundary segregation in {Cu-Bi} alloys using {STEM}.
\newblock \emph{Philos. Mag. A}, {\bfseries {\bfseries 39}\penalty0 (1)},
  \penalty0 1--11.
\newblock \doi{10.1080/01418617908239271}.

\bibitem[{EMD~authors}(2019)]{emd_format}
{EMD~authors}, (2019).
\newblock {Electron Microscopy Datasets: An HDF5-based interchange file format
  for electron microscopy data and metadata}.
\newblock \url{https://emdatasets.com/format}.
\newblock {Accessed} June 3, 2018.

\bibitem[{EMDViewer~devs}(2015)]{emd_viewer}
{EMDViewer~devs}, (2015).
\newblock {EMD Viewer}.
\newblock \url{https://emdatasets.com/viewer}.
\newblock {Accessed} June 3, 2018.

\bibitem[Fang et~al.(2019)Fang, Wen, Allen, Ophus, Han, Kirkland, Kaxiras, and
  Warner]{Fang2019}
Fang, S., Wen, Y., Allen, C.~S., Ophus, C., Han, G. G.~D., Kirkland, A.~I.,
  Kaxiras, E., and Warner, J.~H., (2019).
\newblock Atomic electrostatic maps of {1D} channels in {2D} semiconductors
  using {4D} scanning transmission electron microscopy.
\newblock \emph{Nat. Commun.}, {\bfseries {\bfseries 10}\penalty0 (1)},
  \penalty0 1127.
\newblock \doi{10.1038/s41467-019-08904-9}.

\bibitem[Findlay et~al.(2010)Findlay, Shibata, Sawada, Okunishi, Kondo, and
  Ikuhara]{FINDLAY_2010_UltraMicros_ABF}
Findlay, S.~D., Shibata, N., Sawada, H., Okunishi, E., Kondo, Y., and Ikuhara,
  Y., (2010).
\newblock Dynamics of annular bright field imaging in scanning transmission
  electron microscopy.
\newblock \emph{Ultramicroscopy}, {\bfseries {\bfseries 110}\penalty0 (7)},
  \penalty0 903 -- 923.
\newblock \doi{10.1016/j.ultramic.2010.04.004}.

\bibitem[{fpd~demos~devs}(2018)]{fpd_demos}
{fpd~demos~devs}, (2018).
\newblock {Notebook examples for the fpd package}.
\newblock \url{https://gitlab.com/fpdpy/fpd-demos}.
\newblock {Accessed} June 3, 2018.

\bibitem[{fpd~devs}(2015)]{fpd}
{fpd~devs}, (2015).
\newblock {FPD: Fast pixelated detector data storage, analysis and
  visualisation}.
\newblock \url{https://gitlab.com/fpdpy/fpd}.
\newblock {Accessed} February 6, 2018.

\bibitem[{FPD~Live~Imaging~devs}(2015)]{fpd_live_imaging}
{FPD~Live~Imaging~devs}, (2015).
\newblock {Python library for live visualization of data acquired on a fast
  pixelated detector in a scanning transmission electron microscope}.
\newblock \url{https://gitlab.com/fast_pixelated_detectors/fpd_live_imaging}.
\newblock {Accessed} June 3, 2018.

\bibitem[Frank(1980)]{Frank1980_2imageSNR}
Frank, J.
\newblock (1980), \emph{The Role of Correlation Techniques in Computer Image
  Processing}, pages 187--222.
\newblock Springer Berlin Heidelberg, Berlin, Heidelberg.
\newblock ISBN 978-3-642-81381-8.
\newblock \doi{10.1007/978-3-642-81381-8_5}.

\bibitem[Gouillart et~al.(2016)Gouillart, Nunez-Iglesias, and van~der
  Walt]{Gouillart_2016_python_scikit}
Gouillart, E., Nunez-Iglesias, J., and van~der Walt, S., (2016).
\newblock Analyzing microtomography data with {Python} and the scikit-image
  library.
\newblock \emph{Adv. Struct. Chem. Imaging}, {\bfseries {\bfseries 2}\penalty0
  (1)}, \penalty0 18.
\newblock \doi{10.1186/s40679-016-0031-0}.

\bibitem[{H. Yang} et~al.(2016){H. Yang}, {R. N. Rutte}, {L. Jones}, {M.
  Simson}, {R. Sagawa}, {H. Ryll}, {M. Huth}, {T. J. Pennycook}, {M. L. H.
  Green}, {H. Soltau}, {Y. Kondo}, {B. G. Davis}, and {P. D.
  Nellist}]{Yang_nature_2015_ptychography}
{H. Yang}, {R. N. Rutte}, {L. Jones}, {M. Simson}, {R. Sagawa}, {H. Ryll}, {M.
  Huth}, {T. J. Pennycook}, {M. L. H. Green}, {H. Soltau}, {Y. Kondo}, {B. G.
  Davis}, and {P. D. Nellist}, (2016).
\newblock {Simultaneous atomic-resolution electron ptychography and
  {Z-contrast} imaging of light and heavy elements in complex nanostructures}.
\newblock \emph{Nat. Commun.}, {\bfseries 7}, \penalty0 12532.
\newblock \doi{10.1038/ncomms12532}.

\bibitem[Hachtel et~al.(2018)Hachtel, Idrobo, and Chi]{Hachtel2018}
Hachtel, J.~A., Idrobo, J.~C., and Chi, M., (2018).
\newblock Sub-{\aa}ngstrom electric field measurements on a universal detector
  in a scanning transmission electron microscope.
\newblock \emph{Adv. Struct. Chem. Imaging}, {\bfseries {\bfseries 4}\penalty0
  (1)}, \penalty0 10.
\newblock \doi{10.1186/s40679-018-0059-4}.

\bibitem[Hammel and Rose(1995)]{HAMMEL_1995_UltraMicros_ABF}
Hammel, M. and Rose, H., (1995).
\newblock Optimum rotationally symmetric detector configurations for
  phase-contrast imaging in scanning transmission electron microscopy.
\newblock \emph{Ultramicroscopy}, {\bfseries {\bfseries 58}\penalty0 (3)},
  \penalty0 403 -- 415.
\newblock \doi{10.1016/0304-3991(95)00007-N}.

\bibitem[Hartel et~al.(1996)Hartel, Rose, and
  Dinges]{HARTEL_1996_UltraMicros_haadf}
Hartel, P., Rose, H., and Dinges, C., (1996).
\newblock Conditions and reasons for incoherent imaging in {STEM}.
\newblock \emph{Ultramicroscopy}, {\bfseries {\bfseries 63}\penalty0 (2)},
  \penalty0 93 -- 114.
\newblock \doi{10.1016/0304-3991(96)00020-4}.

\bibitem[House et~al.(2018)House, Tom~Schamp, and
  Yang]{house_schamp_yang_2018_ap_gun_noise}
House, S.~D., Tom~Schamp, C., and Yang, J.~C., (2018).
\newblock {Real Time Acquisition and Calibration of S/TEM Probe Current
  Measurement Simultaneously with Any Imaging or Spectroscopic Signal}.
\newblock \emph{Microsc. Microanal.}, {\bfseries {\bfseries 24}\penalty0 (S1)},
  \penalty0 126–127.
\newblock \doi{10.1017/S1431927618001125}.

\bibitem[Hunter(2007)]{matplotlib}
Hunter, J.~D., (2007).
\newblock Matplotlib: A {2D} graphics environment.
\newblock \emph{Comput. Sci. Eng.}, {\bfseries {\bfseries 9}\penalty0 (3)},
  \penalty0 90--95.
\newblock \doi{10.1109/MCSE.2007.55}.

\bibitem[Huth et~al.(2019)Huth, Ritz, O'Leary, Griffiths, Nellist, and
  Soltau]{huth_2019_MandM_pnccd_fps}
Huth, M., Ritz, R., O'Leary, C.~M., Griffiths, I., Nellist, P., and Soltau, H.,
  (2019).
\newblock {Ultrafast Ptychography with 7500 Frames per Second}.
\newblock \emph{Microsc. Microanal}, {\bfseries {\bfseries 25}\penalty0 (S2)},
  \penalty0 40–41.
\newblock \doi{10.1017/S143192761900093X}.

\bibitem[Johnstone et~al.(2019)Johnstone, Crout, H{\o}g{\aa}s, Martineau,
  Smeets, Laulainen, Collins, Morzy, Prestat, {\AA}nes, phillipcrout, Doherty,
  Ostasevicius, and Bergh]{pyxem}
Johnstone, D.~N., Crout, P., H{\o}g{\aa}s, S., Martineau, B., Smeets, S.,
  Laulainen, J., Collins, S., Morzy, J., Prestat, E., {\AA}nes, H.,
  phillipcrout, Doherty, T., Ostasevicius, T., and Bergh, T., (September 2019).
\newblock pyxem/pyxem: pyxem 0.9.2.
\newblock \url{https://doi.org/10.5281/zenodo.3407316}.

\bibitem[Jones et~al.(2001)Jones, Oliphant, Peterson, et~al.]{scipy}
Jones, E., Oliphant, T., Peterson, P., et~al., (2001).
\newblock {SciPy}: Open source scientific tools for {Python}.
\newblock \url{http://www.scipy.org}.
\newblock {Accessed} 30/10/2018.

\bibitem[Kluyver et~al.(2016)Kluyver, Ragan-Kelley, P{\'e}rez, Granger,
  Bussonnier, Frederic, Kelley, Hamrick, Grout, Corlay, Ivanov, Avila, Abdalla,
  and Willing]{Kluyver_2016_jupyter}
Kluyver, T., Ragan-Kelley, B., P{\'e}rez, F., Granger, B., Bussonnier, M.,
  Frederic, J., Kelley, K., Hamrick, J., Grout, J., Corlay, S., Ivanov, P.,
  Avila, D., Abdalla, S., and Willing, C., (2016).
\newblock Jupyter notebooks -- a publishing format for reproducible
  computational workflows.
\newblock In Loizides, F. and Schmidt, B., editors, \emph{Positioning and Power
  in Academic Publishing: Players, Agents and Agendas}, pages 87 -- 90. IOS
  Press.

\bibitem[K{\"{o}}nnecke et~al.(2015)K{\"{o}}nnecke, Akeroyd, Bernstein,
  Brewster, Campbell, Clausen, Cottrell, Hoffmann, Jemian, M{\"{a}}nnicke,
  Osborn, Peterson, Richter, Suzuki, Watts, Wintersberger, and
  Wuttke]{nexus_format}
K{\"{o}}nnecke, M., Akeroyd, F.~A., Bernstein, H.~J., Brewster, A.~S.,
  Campbell, S.~I., Clausen, B., Cottrell, S., Hoffmann, J.~U., Jemian, P.~R.,
  M{\"{a}}nnicke, D., Osborn, R., Peterson, P.~F., Richter, T., Suzuki, J.,
  Watts, B., Wintersberger, E., and Wuttke, J., (2015).
\newblock {The {NeXus} data format}.
\newblock \emph{J. Appl. Crystallogr.}, {\bfseries {\bfseries 48}\penalty0
  (1)}, \penalty0 301--305.
\newblock \doi{10.1107/S1600576714027575}.

\bibitem[Krajnak et~al.(2016)Krajnak, McGrouther, M., O'Shea, and
  McVitie]{matus_pixelated_stem_magnetic_2016}
Krajnak, M., McGrouther, D., M., D., O'Shea, V., and McVitie, S., (2016).
\newblock Pixelated detectors and improved efficiency for magnetic imaging in
  {STEM} differential phase contrast.
\newblock \emph{Ultramicroscopy}, {\bfseries 165}, \penalty0 42 -- 50.
\newblock \doi{10.1016/j.ultramic.2016.03.006}.

\bibitem[LeBeau et~al.(2009)LeBeau, D'Alfonso, Findlay, Stemmer, and
  Allen]{LeBeau_2009_PRB_BF}
LeBeau, J.~M., D'Alfonso, A.~J., Findlay, S.~D., Stemmer, S., and Allen, L.~J.,
  (2009).
\newblock Quantitative comparisons of contrast in experimental and simulated
  bright-field scanning transmission electron microscopy images.
\newblock \emph{Phys. Rev. B}, {\bfseries 80}, \penalty0 174106.
\newblock \doi{10.1103/PhysRevB.80.174106}.

\bibitem[LeBeau et~al.(2010)LeBeau, Findlay, Allen, and
  Stemmer]{LEBEAU_2010_um_pacbed}
LeBeau, J.~M., Findlay, S.~D., Allen, L.~J., and Stemmer, S., (2010).
\newblock {Position averaged convergent beam electron diffraction: Theory and
  applications}.
\newblock \emph{Ultramicroscopy}, {\bfseries {\bfseries 110}\penalty0 (2)},
  \penalty0 118 -- 125.
\newblock \doi{10.1016/j.ultramic.2009.10.001}.

\bibitem[{LiberTEM~devs}(2018)]{nexus_libertem}
{LiberTEM~devs}, (2018).
\newblock {LiberTEM/nexus-4dstem}.
\newblock \url{https://github.com/LiberTEM/nexus-4dstem}.

\bibitem[MacLaren et~al.(2013)MacLaren, Wang, Morris, Craven, Stamps, Schaffer,
  Ramasse, Miao, Kalantari, Sterianou, and
  Reaney]{MacLaren_2015_APLmat_haadf_bf}
MacLaren, I., Wang, L., Morris, O., Craven, A.~J., Stamps, R.~L., Schaffer, B.,
  Ramasse, Q.~M., Miao, S., Kalantari, K., Sterianou, I., and Reaney, I.~M.,
  (2013).
\newblock {Local stabilisation of polar order at charged antiphase boundaries
  in antiferroelectric (Bi$_{0.85}$Nd$_{0.15}$)(Ti$_{0.1}$Fe$_{0.9}$)O$_3$}.
\newblock \emph{APL Mater.}, {\bfseries {\bfseries 1}\penalty0 (2)}, \penalty0
  021102.
\newblock \doi{10.1063/1.4818002}.

\bibitem[MacLaren et~al.(2020)MacLaren, Frutos-Myro, McGrouther, McFadzean,
  Weiss, Cosart, Portillo, Robins, Nicolopoulos, del Busto, and
  Skogeby]{MacLaren_MandM_2020_sped}
MacLaren, I., Frutos-Myro, E., McGrouther, D., McFadzean, S., Weiss, J.~K.,
  Cosart, D., Portillo, J., Robins, A., Nicolopoulos, S., del Busto, E.~N., and
  Skogeby, R., (2020).
\newblock Orientation mapping using scanned precession electron diffraction
  with a direct electron detector.
\newblock \emph{Microsc. Microanal.}, {\bfseries TBA}, \penalty0 TBA.
\newblock \doi{TBA}.

\bibitem[Mahr et~al.(2019)Mahr, M{\"u}ller-Caspary, Ritz, Simson, Grieb,
  Schowalter, Krause, Lackmann, Soltau, Wittstock, and
  Rosenauer]{knut_muller_nbed_2019}
Mahr, C., M{\"u}ller-Caspary, K., Ritz, R., Simson, M., Grieb, T., Schowalter,
  M., Krause, F.~F., Lackmann, A., Soltau, H., Wittstock, A., and Rosenauer,
  A., (2019).
\newblock Influence of distortions of recorded diffraction patterns on strain
  analysis by nano-beam electron diffraction.
\newblock \emph{Ultramicroscopy}, {\bfseries 196}, \penalty0 74 -- 82.
\newblock \doi{10.1016/j.ultramic.2018.09.010}.

\bibitem[McGrouther et~al.(2014)McGrouther, Benitez~Romero, McFadzean, and
  McVitie]{mcgrouther2014development}
McGrouther, D., Benitez~Romero, M.~J., McFadzean, S., and McVitie, S., (2014).
\newblock Development of aberration corrected differential phase contrast
  {(DPC) STEM}.
\newblock \emph{JEOL News}, {\bfseries 49}\penalty0 (1).

\bibitem[McGrouther et~al.(2015)McGrouther, Krajnak, MacLaren, Maneuski,
  O'Shea, and Nellist]{mcgrouther_medipix_stem_2015}
McGrouther, D., Krajnak, M., MacLaren, I., Maneuski, D., O'Shea, V., and
  Nellist, P.~D., (2015).
\newblock Use of a hybrid silicon pixel ({Medipix}) detector as a {STEM}
  detector.
\newblock \emph{Microsc. Microanal.}, {\bfseries 21}, \penalty0 1595.
\newblock \doi{10.1017/S1431927615008752}.

\bibitem[McMullan et~al.(2007)McMullan, Cattermole, Chen, Henderson, Llopart,
  Summerfield, Tlustos, and Faruqi]{MCMULLAN_UM_2007_mpx_high_z_comment}
McMullan, G., Cattermole, D.~M., Chen, S., Henderson, R., Llopart, X.,
  Summerfield, C., Tlustos, L., and Faruqi, A.~R., (2007).
\newblock Electron imaging with {Medipix2} hybrid pixel detector.
\newblock \emph{Ultramicroscopy}, {\bfseries {\bfseries 107}\penalty0 (4)},
  \penalty0 401 -- 413.
\newblock \doi{10.1016/j.ultramic.2006.10.005}.

\bibitem[McMullan et~al.(2014)McMullan, Faruqi, Clare, and
  Henderson]{DEDComparison}
McMullan, G., Faruqi, A.~R., Clare, D., and Henderson, R., (2014).
\newblock Comparison of optimal performance at 300 {keV} of three direct
  electron detectors for use in low dose electron microscopy.
\newblock \emph{Ultramicroscopy}, {\bfseries 147}, \penalty0 156--163.
\newblock \doi{10.1016/j.ultramic.2014.08.002}.

\bibitem[{Merlin~DM~Plugin~devs}(2017)]{merlin_dm_plugin}
{Merlin~DM~Plugin~devs}, (2017).
\newblock {Digital Micrograph scripts and plugin for communicating between DM
  and the Merlin Medipix readout and control software, potentially allowing
  full control of the Medipix detector}.
\newblock \url{https://gitlab.com/fast_pixelated_detectors/merlin_dm_plugin}.
\newblock {Accessed} October 14, 2019.

\bibitem[{Merlin~Interface~devs}(2016)]{merlin_interface}
{Merlin~Interface~devs}, (2016).
\newblock {Python library for interfacing with a {Medipix3} detector through
  the {Merlin} software through the TCP/IP command interface}.
\newblock \url{https://gitlab.com/fast_pixelated_detectors/merlin_interface}.
\newblock {Accessed} June 3, 2018.

\bibitem[Midgley and Eggeman(2015)]{Midgley_IUCrJ_2015_ped_rev}
Midgley, P.~A. and Eggeman, A.~S., (2015).
\newblock {Precession electron diffraction {--} a topical review}.
\newblock \emph{IUCrJ}, {\bfseries {\bfseries 2}\penalty0 (1)}, \penalty0
  126--136.
\newblock \doi{10.1107/S2052252514022283}.

\bibitem[Mir et~al.(2017)Mir, Clough, MacInnes, Gough, Plackett, Shipsey,
  Sawada, MacLaren, Ballabriga, Maneuski, O'Shea, McGrouther, and
  Kirkland]{MIR_2017_medipix3_characterisation}
Mir, J.~A., Clough, R., MacInnes, R., Gough, C., Plackett, R., Shipsey, I.,
  Sawada, H., MacLaren, I., Ballabriga, R., Maneuski, D., O'Shea, V.,
  McGrouther, D., and Kirkland, A.~I., (2017).
\newblock Characterisation of the {Medipix3} detector for 60 and {80keV}
  electrons.
\newblock \emph{Ultramicroscopy}, {\bfseries 182}, \penalty0 44 -- 53.
\newblock \doi{10.1016/j.ultramic.2017.06.010}.

\bibitem[{Niermann, T}(2016)]{DM_hdf5}
{Niermann, T}, (2016).
\newblock Digital {M}icrograph plugin to read/write {HDF5} files.
\newblock \url{https://github.com/niermann/gms_plugin_hdf5}.
\newblock {Accessed} June 3, 2018.

\bibitem[Nord et~al.(2016)Nord, Krajnak, Bali, Hlawacek, Liersch, Fassbender,
  McVitie, Paterson, Maclaren, and McGrouther]{nord_magnetic_2016}
Nord, M., Krajnak, M., Bali, R., Hlawacek, G., Liersch, V., Fassbender, J.,
  McVitie, S., Paterson, G.~W., Maclaren, I., and McGrouther, D., (2016).
\newblock Developing rapid and advanced visualisation of magnetic structures
  using {2-D} pixelated {STEM} detectors.
\newblock \emph{Microsc. Microanal.}, {\bfseries {\bfseries 22}\penalty0 (S3)},
  \penalty0 530--531.
\newblock \doi{10.1017/S1431927616003500}.

\bibitem[Nord et~al.(2017)Nord, Vullum, MacLaren, Tybell, and
  Holmestad]{Nord2017_atomap}
Nord, M., Vullum, P.~E., MacLaren, I., Tybell, T., and Holmestad, R., (2017).
\newblock Atomap: a new software tool for the automated analysis of atomic
  resolution images using two-dimensional {G}aussian fitting.
\newblock \emph{Adv. Struct. Chem. Imaging}, {\bfseries {\bfseries 3}\penalty0
  (1)}, \penalty0 9.
\newblock \doi{10.1186/s40679-017-0042-5}.

\bibitem[Nord et~al.(2019{\natexlab{a}})Nord, Ross, McGrouther, Barthel,
  Moreau, Hallsteinsen, Tybell, and MacLaren]{Nord_2018_holz}
Nord, M., Ross, A., McGrouther, D., Barthel, J., Moreau, M., Hallsteinsen, I.,
  Tybell, T., and MacLaren, I., (2019{\natexlab{a}}).
\newblock Three-dimensional subnanoscale imaging of unit cell doubling due to
  octahedral tilting and cation modulation in strained perovskite thin films.
\newblock \emph{Phys. Rev. Mater.}, {\bfseries 3}, \penalty0 063605.
\newblock \doi{10.1103/PhysRevMaterials.3.063605}.

\bibitem[Nord et~al.(2019{\natexlab{b}})Nord, Webster, Paton, McVitie,
  McGrouther, MacLaren, and Paterson]{paper_dataset_zenodo}
Nord, M., Webster, R. W.~H., Paton, K.~A., McVitie, S., McGrouther, D.,
  MacLaren, I., and Paterson, G.~W., (2019{\natexlab{b}}).
\newblock {Dataset}.
\newblock \url{https://doi.org/10.5281/zenodo.3479124}.

\bibitem[Oliphant(2006)]{numpy}
Oliphant, T.~E., (2006).
\newblock \emph{A guide to {NumPy}}.
\newblock USA: Trelgol Publishing.

\bibitem[Oliphant(2007)]{oliphant_07_python}
Oliphant, T.~E., (2007).
\newblock Python for scientific computing.
\newblock \emph{Comput. Sci. Eng.}, {\bfseries {\bfseries 9}\penalty0 (3)},
  \penalty0 10--20.
\newblock \doi{10.1109/MCSE.2007.58}.

\bibitem[Ophus(2019)]{ophus_mm_2019_4dstem}
Ophus, C., (2019).
\newblock {Four-Dimensional Scanning Transmission Electron Microscopy
  (4D-STEM): From Scanning Nanodiffraction to Ptychography and Beyond}.
\newblock \emph{Microsc. Microanal.}, {\bfseries {\bfseries 25}\penalty0 (3)},
  \penalty0 563–582.
\newblock \doi{10.1017/S1431927619000497}.

\bibitem[Ophus et~al.(2017)Ophus, Ercius, Huijben, and
  Ciston]{Ophus_APL_2017_pacbed}
Ophus, C., Ercius, P., Huijben, M., and Ciston, J., (2017).
\newblock Non-spectroscopic composition measurements of
  {SrTiO$_{3}$-La$_{0.7}$Sr$_{0.3}$MnO$_3$} multilayers using scanning
  convergent beam electron diffraction.
\newblock \emph{Appl. Phys. Lett.}, {\bfseries {\bfseries 110}\penalty0 (6)},
  \penalty0 063102.
\newblock \doi{10.1063/1.4975932}.

\bibitem[Ophus(2017)]{prismatic}
Ophus, C., (May 2017).
\newblock A fast image simulation algorithm for scanning transmission electron
  microscopy.
\newblock \emph{Adv. Struct. Chem. Imaging}, {\bfseries {\bfseries 3}\penalty0
  (1)}, \penalty0 13.
\newblock \doi{10.1186/s40679-017-0046-1}.

\bibitem[Paterson et~al.(2020{\natexlab{a}})Paterson, Lamb, Ballabriga,
  Maneuski, O'Shea, and McGrouther]{medipix3_time_resolution}
Paterson, G.~W., Lamb, R.~J., Ballabriga, R., Maneuski, D., O'Shea, V., and
  McGrouther, D., (2020{\natexlab{a}}).
\newblock {Sub-100 nanosecond temporally resolved imaging with the Medipix3
  direct electron detector}.
\newblock \emph{Ultramicroscopy}, {\bfseries 210}, \penalty0 112917.
\newblock \doi{10.1016/j.ultramic.2019.112917}.

\bibitem[Paterson et~al.(2020{\natexlab{b}})Paterson, Webster, Ross, Paton,
  Macgregor, McGrouther, MacLaren, and Nord]{methods_part2_arxiv}
Paterson, G.~W., Webster, R. W.~H., Ross, A., Paton, K.~A., Macgregor, T.~A.,
  McGrouther, D., MacLaren, I., and Nord, M., (2020{\natexlab{b}}).
\newblock {Fast Pixelated Detectors in Scanning Transmission Electron
  Microscopy. Part II: Post Acquisition Data Processing, Visualisation, and
  Structural Characterisation}.
\newblock \url{arXiv:2004.02777}.

\bibitem[Pennycook and Jesson(1991)]{Pennycook1991_um}
Pennycook, S. and Jesson, D., (1991).
\newblock High-resolution {Z}-contrast imaging of crystals.
\newblock \emph{Ultramicroscopy}, {\bfseries {\bfseries 37}\penalty0 (1)},
  \penalty0 14 -- 38.
\newblock \doi{10.1016/0304-3991(91)90004-P}.

\bibitem[Pennycook et~al.(2015)Pennycook, Lupini, Yang, Murfitt, Jones, and
  Nellist]{pennycook_ptychography_2015}
Pennycook, T.~J., Lupini, A.~R., Yang, H., Murfitt, M.~F., Jones, L., and
  Nellist, P.~D., (2015).
\newblock {Efficient phase contrast imaging in STEM using a pixelated detector.
  Part 1: Experimental demonstration at atomic resolution}.
\newblock \emph{Ultramicroscopy}, {\bfseries 151}, \penalty0 160--167.
\newblock \doi{10.1016/j.ultramic.2014.09.013}.

\bibitem[{pixStem~devs}(2015)]{pixstem}
{pixStem~devs}, (2015).
\newblock {pixStem: Analysis of pixelated {STEM} data}.
\newblock \url{https://gitlab.com/pixstem/pixstem}.
\newblock {Accessed} October 3, 2018.

\bibitem[Plackett et~al.(2013)Plackett, Horswell, Gimenez, Marchal, Omar, and
  Tartoni]{Plackett_2013_merlin}
Plackett, R., Horswell, I., Gimenez, E.~N., Marchal, J., Omar, D., and Tartoni,
  N., (2013).
\newblock Merlin: a fast versatile readout system for {Medipix3}.
\newblock \emph{J. Instrum.}, {\bfseries {\bfseries 8}\penalty0 (01)},
  \penalty0 C01038.
\newblock \doi{10.1088/1748-0221/8/01/C01038}.

\bibitem[Poikela et~al.(2014)Poikela, Plosila, Westerlund, Campbell, Gaspari,
  Llopart, Gromov, Kluit, van Beuzekom, Zappon, Zivkovic, Brezina, Desch, Fu,
  and Kruth]{Poikela_2014_JoI_Timepix3}
Poikela, T., Plosila, J., Westerlund, T., Campbell, M., Gaspari, M.~D.,
  Llopart, X., Gromov, V., Kluit, R., van Beuzekom, M., Zappon, F., Zivkovic,
  V., Brezina, C., Desch, K., Fu, Y., and Kruth, A., (2014).
\newblock {Timepix3: a 65K channel hybrid pixel readout chip with simultaneous
  {ToA}/{ToT} and sparse readout}.
\newblock \emph{J. Instrum.}, {\bfseries {\bfseries 9}\penalty0 (05)},
  \penalty0 C05013--C05013.
\newblock \doi{10.1088/1748-0221/9/05/c05013}.

\bibitem[Rauch et~al.(2010)Rauch, Portillo, Nicolopoulos, Bultreys, Rouvimov,
  and Moeck]{Rauch_SPED_2010}
Rauch, E.~F., Portillo, J., Nicolopoulos, S., Bultreys, D., Rouvimov, S., and
  Moeck, P., (2010).
\newblock Automated nanocrystal orientation and phase mapping in the
  transmission electron microscope on the basis of precession electron
  diffraction.
\newblock \emph{Z. Kristallogr.}, {\bfseries 225}, \penalty0 103--109.
\newblock \doi{10.1524/zkri.2010.1205}.

\bibitem[Ryll et~al.(2016)Ryll, Simson, Hartmann, Holl, Huth, Ihle, Kondo,
  Kotula, Liebel, Müller-Caspary, Rosenauer, Sagawa, Schmidt, Soltau, and
  Strüder]{Ryll_2016_pnccd}
Ryll, H., Simson, M., Hartmann, R., Holl, P., Huth, M., Ihle, S., Kondo, Y.,
  Kotula, P., Liebel, A., Müller-Caspary, K., Rosenauer, A., Sagawa, R.,
  Schmidt, J., Soltau, H., and Strüder, L., (2016).
\newblock {A pnCCD-based, fast direct single electron imaging camera for TEM
  and STEM}.
\newblock \emph{J. Instrum.}, {\bfseries {\bfseries 11}\penalty0 (04)},
  \penalty0 P04006--P04006.
\newblock \doi{10.1088/1748-0221/11/04/p04006}.

\bibitem[Savitzky et~al.(2018)Savitzky, Baggari, Clement, Waite, Goodge, Baek,
  Sheckelton, Pasco, Nair, Schreiber, Hoffman, Admasu, Kim, Cheong,
  Bhattacharya, Schlom, McQueen, Hovden, and Kourkoutis]{rigid_registration}
Savitzky, B.~H., Baggari, I.~E., Clement, C.~B., Waite, E., Goodge, B.~H.,
  Baek, D.~J., Sheckelton, J.~P., Pasco, C., Nair, H., Schreiber, N.~J.,
  Hoffman, J., Admasu, A.~S., Kim, J., Cheong, S.-W., Bhattacharya, A., Schlom,
  D.~G., McQueen, T.~M., Hovden, R., and Kourkoutis, L.~F., (2018).
\newblock {Image registration of low signal-to-noise cryo-STEM data}.
\newblock \emph{Ultramicroscopy}, {\bfseries 191}, \penalty0 56 -- 65.
\newblock \doi{10.1016/j.ultramic.2018.04.008}.

\bibitem[Savitzky et~al.(2019)Savitzky, Zeltmann, Barnard, lerandc, Brown,
  Henderson, and Ginsburg]{py4dstem}
Savitzky, B.~H., Zeltmann, S., Barnard, E., lerandc, Brown, H.~G., Henderson,
  M., and Ginsburg, D., (July 2019).
\newblock {py4dstem/py4DSTEM: DOI release}.
\newblock \url{https://doi.org/10.5281/zenodo.3333960}.

\bibitem[Scholze et~al.(1998)Scholze, Rabus, and Ulm]{scholze_mean_1998}
Scholze, F., Rabus, H., and Ulm, G., (1998).
\newblock Mean energy required to produce an electron-hole pair in silicon for
  photons of energies between 50 and 1500 {eV}.
\newblock \emph{J. of Appl. Phys.}, {\bfseries {\bfseries 84}\penalty0 (5)},
  \penalty0 2926--2939.
\newblock \doi{10.1063/1.368398}.

\bibitem[Shibata et~al.(2017)Shibata, Seki, S{\'a}nchez-Santolino, Findlay,
  Kohno, Matsumoto, Ishikawa, and
  Ikuhara]{Shibata2017_nature_atom_electric_fields}
Shibata, N., Seki, T., S{\'a}nchez-Santolino, G., Findlay, S.~D., Kohno, Y.,
  Matsumoto, T., Ishikawa, R., and Ikuhara, Y., (2017).
\newblock Electric field imaging of single atoms.
\newblock \emph{Nat. Commun.}, {\bfseries 8}, \penalty0 15631.
\newblock \doi{10.1038/ncomms15631}.

\bibitem[Shibata et~al.(2010)Shibata, Kohno, Findlay, Sawada, Kondo, and
  Ikuhara]{Shibata2010_detector}
Shibata, N., Kohno, Y., Findlay, S.~D., Sawada, H., Kondo, Y., and Ikuhara, Y.,
  (04 2010).
\newblock {New area detector for atomic-resolution scanning transmission
  electron microscopy}.
\newblock \emph{Microscopy}, {\bfseries {\bfseries 59}\penalty0 (6)}, \penalty0
  473--479.
\newblock \doi{10.1093/jmicro/dfq014}.

\bibitem[Somnath et~al.(2019)Somnath, Smith, Laanait, Vasudevan, Ievlev,
  Belianinov, Lupini, Shankar, Kalinin, and Jesse]{pycroscopy}
Somnath, S., Smith, C.~R., Laanait, N., Vasudevan, R.~K., Ievlev, A.,
  Belianinov, A., Lupini, A.~R., Shankar, M., Kalinin, S.~V., and Jesse, S.,
  (2019).
\newblock {USID and Pycroscopy -- Open frameworks for storing and analyzing
  spectroscopic and imaging data}.
\newblock \url{arXiv:1903.09515}.

\bibitem[Taguchi and Iwanczyk(2013)]{Taguchi_2013_ct}
Taguchi, K. and Iwanczyk, J.~S., (2013).
\newblock {Vision 20/20: Single photon counting x-ray detectors in medical
  imaging}.
\newblock \emph{Med. Phys.}, {\bfseries 40}, \penalty0 100901.
\newblock \doi{10.1118/1.4820371}.

\bibitem[Tate et~al.(2016)Tate, Purohit, Chamberlain, Nguyen, Hovden, Chang,
  Deb, Turgut, Heron, Schlom, Ralph, Fuchs, Shanks, Philipp, Muller, and
  Gruner]{tate_2016_mandm_empad}
Tate, M.~W., Purohit, P., Chamberlain, D., Nguyen, K.~X., Hovden, R., Chang,
  C.~S., Deb, P., Turgut, E., Heron, J.~T., Schlom, D.~G., Ralph, D.~C., Fuchs,
  G.~D., Shanks, K.~S., Philipp, H.~T., Muller, D.~A., and Gruner, S.~M.,
  (2016).
\newblock High dynamic range pixel array detector for scanning transmission
  electron microscopy.
\newblock \emph{Microsc. Microanal.}, {\bfseries {\bfseries 22}\penalty0 (1)},
  \penalty0 237–249.
\newblock \doi{10.1017/S1431927615015664}.

\bibitem[{The HDF Group}(1997-2018)]{hdf5_file_format}
{The HDF Group}, (1997-2018).
\newblock {Hierarchical Data Format, version 5}.
\newblock \url{http://www.hdfgroup.org/HDF5}.

\bibitem[{The~HDF~Group}(2017)]{HDFView}
{The~HDF~Group}, (2017).
\newblock {HDFView is a visual tool for browsing and editing HDF4 and HDF5
  files}.
\newblock \url{https://support.hdfgroup.org/products/java/hdfview}.
\newblock {Accessed} June 3, 2018.

\bibitem[Tinti et~al.(2018)Tinti, Fr{\"{o}}jdh, van Genderen, Gruene, Schmitt,
  de~Winter, Weckhuysen, and Abrahams]{Tinti_2018_eiger}
Tinti, G., Fr{\"{o}}jdh, E., van Genderen, E., Gruene, T., Schmitt, B.,
  de~Winter, D. A.~M., Weckhuysen, B.~M., and Abrahams, J.~P., (2018).
\newblock {Electron crystallography with the {EIGER} detector}.
\newblock \emph{IUCrJ}, {\bfseries {\bfseries 5}\penalty0 (2)}, \penalty0
  190--199.
\newblock \doi{10.1107/S2052252518000945}.

\bibitem[Turala(2005)]{Turala_2005_SI_tracking}
Turala, M., (2005).
\newblock Silicon tracking detectors-historical overview.
\newblock \emph{Nucl. Instrum. Methods Phys. Res. A}, 541.
\newblock \doi{10.1016/j.nima.2005.01.032}.

\bibitem[Turchetta et~al.(2007)Turchetta, Fant, Gasiorek, Esbrand, Griffiths,
  Metaxas, Royle, Speller, Venanzi, van~der Stelt, Verheij, Li, Theodoridis,
  Georgiou, Cavouras, Hall, Noy, Jones, Leaver, Machin, Greenwood, Khaleeq,
  Schulerud, {\O}stby, Triantis, Asimidis, Bolanakis, Manthos, Longo, and
  Bergamaschi]{Turchetta2007_maps}
Turchetta, R., Fant, A., Gasiorek, P., Esbrand, C., Griffiths, J., Metaxas, M.,
  Royle, G., Speller, R., Venanzi, C., van~der Stelt, P., Verheij, H., Li, G.,
  Theodoridis, S., Georgiou, H., Cavouras, D., Hall, G., Noy, M., Jones, J.,
  Leaver, J., Machin, D., Greenwood, S., Khaleeq, M., Schulerud, H., {\O}stby,
  J., Triantis, F., Asimidis, A., Bolanakis, D., Manthos, N., Longo, R., and
  Bergamaschi, A., (2007).
\newblock {CMOS Monolithic Active Pixel Sensors (MAPS): Developments and future
  outlook}.
\newblock \emph{Nucl. Instrum. Methods Phys. Res. A}, {\bfseries {\bfseries
  582}\penalty0 (3)}, \penalty0 866 -- 870.
\newblock \doi{10.1016/j.nima.2007.07.112}.

\bibitem[van~der Walt et~al.(2014)van~der Walt, {S}ch\"onberger,
  {Nunez-Iglesias}, {B}oulogne, {W}arner, {Y}ager, {G}ouillart, {Y}u, and the
  scikit-image contributors]{scikit_image}
van~der Walt, S., {S}ch\"onberger, J.~L., {Nunez-Iglesias}, J., {B}oulogne, F.,
  {W}arner, J.~D., {Y}ager, N., {G}ouillart, E., {Y}u, T., and the scikit-image
  contributors, (2014).
\newblock scikit-image: image processing in {P}ython.
\newblock \emph{PeerJ}, {\bfseries 2}, \penalty0 e453.
\newblock \doi{10.7717/peerj.453}.

\bibitem[von Ardenne(1938)]{von_Ardenne_1938_tem_theory}
von Ardenne, M., (1938).
\newblock The scanning electron microscope: {Theoretical} fundamentals.
\newblock \emph{Z. Phys.}, {\bfseries 109}, \penalty0 553–572.
\newblock \doi{10.1007/BF01341584}.

\bibitem[Wermes(2005)]{Wermes_2005_HPD_mono}
Wermes, N., (2005).
\newblock Pixel detectors for tracking and their spin-off in imaging
  applications.
\newblock \emph{Nucl. Instrum. Methods Phys. Res. A}, {\bfseries {\bfseries
  541}\penalty0 (1)}, \penalty0 150 -- 165.
\newblock \doi{10.1016/j.nima.2005.01.052}.

\end{thebibliography}

\onecolumngrid

\end{document}